\documentclass[a4paper,11pt]{article}
\usepackage{jheppub}
\usepackage{amsthm}
\usepackage[T1]{fontenc}
\usepackage[utf8]{inputenc}
\usepackage{lmodern}
\usepackage{microtype}
\title{\boldmath {Wald entropy and Wen-Zee type gauge-curvature coupling in 2+1-dimensional gravity}}

\author[a]{Aryan Bhatia,}
\author[b]{Giandomenico Palumbo,}
\author[a]{Suraj S. Hegde}
\affiliation[a]{Indian Institute of Science Education and Research Thiruvananthapuram, Vithura, Thiruvananthapuram, Kerala, India}
\affiliation[b]{CFisUC, Department of Physics, University of Coimbra, Rua Larga, 3004-516 Coimbra, Portugal}

\emailAdd{surajhegde@iisertvm.ac.in}

\abstract{The Wen-Zee term is a mixed gauge-curvature coupling in the topological Chern-Simons description of a quantum Hall fluid, encoding the electric charge bound to spatial curvature. We investigate a Lorentz-covariant generalization of this coupling, expanding its scope to black-hole spacetimes and determine its contribution to horizon entropy. Such a coupling arises naturally in three-dimensional Chern-Simons gravity based on the Maxwell algebra. We show that its Wald entropy contribution is a horizon boost charge, given by a loop integral of the additional gauge field projected along the tangent to the bifurcation circle. This contribution can be nonzero even when the mixed term vanishes on shell. We show that the integral admits an Abelian holonomy interpretation when the projection direction is covariantly constant. We evaluate the charge on a flat cosmological solution and, using smooth horizon fields and a specified thermal ensemble, on an AdS-Lorentz black hole. These results connect the geometric response familiar from quantum Hall physics to gravitational horizon thermodynamics.}

\begin{document}
\maketitle
\flushbottom

\section{Introduction}
\label{sec:intro}

Topological quantum field theories (TQFTs) provide a common language for describing physical phenomena across diverse topics such as condensed matter physics \cite{Wen2004} and gravity \cite{Birmingham1991}. Among TQFTs we are particularly interested in Chern-Simons theories in $(2+1)$-dimensional spacetime \cite{Dunne1999}. In condensed matter systems, they capture the universal low-energy properties of topological phases, particularly quantum Hall phases both in the continuum \cite{Zhang1989,Lopez1991,Fradkin1998} and lattice systems \cite{Yakovenko1990,Palumbo2013,Palumbo2014,Salgado2020}. For the phases they describe, these theories encode fractional charges, quasiparticle statistics and topological ground-state degeneracy~\cite{Wen1995}.  On the other hand, Witten's work established a fundamental relation between gauge invariant objects in TQFTs, namely Wilson loops, and knot invariants~\cite{Witten1989}. The gauge formulation of three-dimensional gravity provides another important application of TQFTs~\cite{AchucarroTownsend1986,Witten1988}.

In the context of quantum Hall physics, an interesting addition to the low-energy description arises when quantum Hall phases are coupled to a background curved manifold. The effective action contains a mixed coupling between the electromagnetic potential and the spatial spin connection, known as the Wen-Zee term~\cite{WenZee1992}. Its coefficient encodes the average orbital spin $\bar s$ and, in the standard convention, the shift $\mathcal S=2\bar s$. This coupling describes the electric charge induced by spatial curvature and relates geometric response to the non-dissipative Hall viscosity of the fluid~\cite{Read2009,BradlynRead2015}. In particular, for an isotropic gapped quantum Hall fluid in the appropriate low-frequency limit, $\eta_H=\hbar\bar n\bar s/2$, where $\bar n$ is the mean particle density.  There have been many studies of these geometric responses in non-relativistic settings~\cite{Hoyos2012,AbanovGromov2014,Can2014, BradlynRead2015,GromovFramingAnomaly2015,Bradlyn2015,Can2016,CappelliRandellini2016, Han2019, Rao2023}. Relativistic constructions have also been developed, using either a conserved Euler current~\cite{GolkarRobertsSon2014,GolkarRobertsSonEuler2015} or a Chern-Simons gauge theory based on the Maxwell extension of the Poincar\'e algebra~\cite{Palumbo2016}. These constructions require additional structure because the spatial $SO(2)$ spin connection of the non-relativistic theory is replaced by the full Lorentz $SO(2,1)$ spin connection in a relativistic setting.

On the other hand, there has been considerable interest in the formulation of gravitational theories in odd spacetime dimensions as gauge theories~\cite{Zanelli2012}. In particular, pure Einstein gravity in $(2+1)$ dimensions can be written in first-order form as a Chern-Simons action~\cite{AchucarroTownsend1986,Witten1988}.
Despite the absence of local propagating gravitational degrees of freedom, the theory when equipped with a negative cosmological constant admits the BTZ black hole~\cite{BTZ1992}. Maxwell extensions of the gauge algebra provide a broader class of three-dimensional Chern-Simons gravitational theories~\cite{Concha2018,Salgado2019,ToplogicalgravityTH,2025boundarydynamicsmaxwellinvariantthreedimensional}  including cosmological solutions and black holes with torsion \cite{Aviles2026}. Of particular relevance here, the additional gauge fields allow mixed gauge-curvature couplings of the Wen-Zee type~\cite{Palumbo2016,DurkaKG2019}, providing a connection between the geometric response of quantum Hall systems and the gauge formulation of gravity.

The scope of such Wen-Zee-type terms can therefore be expanded beyond their relevance to quantum Hall phenomena. One can study them as Lorentz-covariant curvature couplings in gauge theories on curved
spacetimes. We focus on a coupling of the form
$I_{\mathrm{mix}}=\beta\int A_a\wedge R^a$, where $R^a$ is the dual Lorentz curvature, $A^a$ is a Lorentz-vector-valued one-form belonging to
the additional gauge sector, and $\beta$ is the coupling coefficient. The field $A^a$ is distinct from the ordinary electromagnetic potential;
an electromagnetic interpretation requires a further physical identification. Within this expanded scope, we pose the question: what is the contribution of such a gauge-curvature coupling to Wald entropy
in the presence of a horizon?

Black-hole entropy is a central object in the study of gravitational thermodynamics~\cite{Bekenstein1973,Hawking1975}. In a general diffeomorphism-invariant theory, the Wald construction relates it to a horizon Noether charge and makes its dependence on curvature couplings in the action explicit~\cite{Wald1993,IyerWald1994}. In a first-order description, the relevant symmetry combines a diffeomorphism with a local Lorentz transformation~\cite{JacobsonMohd}. Applying this framework to the mixed gauge-curvature coupling, we identify its contribution to the horizon Lorentz-boost charge as a loop integral of the gauge field projected along the dual binormal at the bifurcation circle, where the horizon-generating Killing vector vanishes. In the Maxwell sector, when this projection direction is covariantly constant along the circle, the integral admits a reduced Abelian holonomy interpretation under single-valued gauge shifts. Holonomies also play an important role in the thermodynamics of three-dimensional Chern-Simons gravity, where regularity constrains the thermal holonomy and black-hole entropy can be expressed in terms of the gauge connections~\cite{BunsterEtAl2014,deBoerJottar2014}.

Our work thus examines how the gauge-curvature structure familiar from quantum Hall physics appears in horizon entropy, and specifies the conditions under which this entropy contribution can be understood in terms of a horizon holonomy.

We evaluate the charge on two backgrounds. For the flat cosmological solution of Maxwell Chern-Simons gravity, a smooth representative and the geometric Lorentz lift reproduce the known horizon entropy~\cite{Aviles2026}. For the AdS-Lorentz black hole~\cite{AvilesHidalgoValdivia2023}, we construct a representative with a regular dreibein and independent connection at the outer bifurcation circle. The construction preserves the BTZ metric and the exterior gauge orbit while changing the contorsion. Selecting the lift that becomes the geometric boost at the bifurcation circle gives the unit thermal normalization and allows the full entropy to be computed there, once a discrete independent-sector branch is fixed. The mixed-term projection then reproduces the complete $\alpha_2$ block in this representative; its identification is not a separate invariant of the full AdS-Lorentz gauge group.

\section{Relativistic Wen-Zee term}
\label{sec:RelWenZee}

We start this section by first recapitulating the low-energy limit description of non-relativistic Abelian fractional quantum Hall states that is encoded by the following topological effective field theory in curved space \cite{WenZee1992, AbanovGromov2014, SalgadoRebolledoPalumbo2021, GromovFramingAnomaly2015}

\begin{equation}
 S_{\text{eff}}=\frac{\nu}{4\pi}\int_{\mathcal M} \Big(A\wedge dA+2\bar sA\wedge d\omega\Big)+\left(\frac{\nu\overline{ s^{2}}}{4\pi}-\frac{c}{48\pi}\right)\int_{\mathcal M}\omega\wedge d\omega.
\label{eq:Seff}
\end{equation}
Here $A$ is the electromagnetic potential, $\omega$ the spin connection, $\nu$ the filling factor and $\bar{s}$ the average orbital spin. For an Abelian quantum Hall state characterized by a nonsingular $K$-matrix, charge vector $t$ and orbital-spin vector $s$, the coefficients in the above action are:  filling $\nu=t^T K^{-1}t $, $\nu\bar s=t^T K^{-1}s$ and $\nu \overline{s^2}=s^T K^{-1}s$.
The first term is the Abelian Chern-Simons term which gives the Hall conductance. The second term, known as the first Wen-Zee term, represents a topological coupling between the electromagnetic and geometric sectors. The last term combines the second Wen-Zee coupling, proportional to $\nu\overline{s^2}$, with the anomaly contribution proportional to the chiral central charge $c$. These coefficients encode distinct geometric data.
Varying \eqref{eq:Seff} with respect to $A_{0}$, we obtain
\begin{equation}
\rho_{\rm charge} \;=\; \frac{\nu}{2\pi}B \;+\; \frac{\nu\bar{s}}{4\pi}\,\mathcal{R},
\label{eq:boundcharge}
\end{equation}
where $\mathcal{R}= 2\varepsilon^{ij}\partial_i\omega_j$, with $\varepsilon^{ij}\equiv\frac{1}{|e|}\epsilon^{ij}$.
Integrating \eqref{eq:boundcharge} over a closed surface and using the Gauss-Bonnet theorem gives the shift relation for the total number of electrons $N_e=\nu N_{\phi} \;+\; \nu\bar{s}\,\chi,$
where $N_{\phi}$ is total number of magnetic-flux quanta piercing the closed spatial surface and $\chi$ is the Euler characteristic of the surface~\cite{WenZee1992}.

One important point to note is that $\omega$ is the SO(2) spatial spin connection and encodes the non-relativistic curved-space background. We remind the reader that, although quantum Hall phases in the continuum are intrinsically non-relativistic, Landau levels, Hall conductivity, and topological phases can also arise in lattice models that support relativistic quasiparticles, such as Dirac materials \cite{Novoselov2005,Zhang2005,GusyninSharapov2005}. It is therefore natural to ask whether there also exists a relativistic version of the (first) Wen-Zee term \cite{GolkarRobertsSon2014,GolkarRobertsSonEuler2015}. We will discuss two possible approaches below.

The non-relativistic spin connection is Abelian and for this reason the Wen-Zee term coincides with a mixed Abelian Chern-Simons term. 
Its direct transcription to a relativistic \((2+1)\)-dimensional
setting is, however, not immediate. The Lorentz spin connection is an
\(SO(1,2)\) connection and is therefore non-Abelian, with no obvious coupling with the electromagnetic gauge potential. Two complementary relativistic
constructions have consequently been proposed. The first, developed in
Ref.~\cite{GolkarRobertsSon2014}, uses a conserved Euler current built
from a distinguished unit timelike vector field. The second formulation based on gauging the Maxwell algebra \cite{Palumbo2016, DurkaKG2019},
to be discussed below, is naturally expressed in a first-order geometric
language and makes the local Lorentz structure more explicit.

\subsection{Euler-Current-Curvature coupling}
Following \cite{GolkarRobertsSon2014, GolkarRobertsSonEuler2015}, consider an oriented
Lorentzian three-manifold \((\mathcal M,g_{\mu\nu})\) endowed with a smooth,
nowhere-vanishing vector field \(u^\mu\) of fixed norm
\begin{equation}
u^\mu u_\mu=\sigma,\qquad \sigma=\pm1.
\end{equation}
Throughout we use the metric signature \((-,+,+)\) and take
\(u^\mu\) to be timelike, so that \(u^\mu u_\mu=-1\), i.e.\ \(\sigma=-1\).
Thus \(\sigma\) is fixed by the signature convention together with the timelike character of \(u^\mu\) and is not an
independent coupling. The Euler current is
\begin{equation}
J^\mu=\frac{1}{8\pi}\,
\varepsilon^{\mu\nu\rho}\varepsilon^{\alpha\beta\gamma}u_\alpha
\left(
\nabla_\nu u_\beta\,\nabla_\rho u_\gamma
-\frac{R_{\nu\rho\beta\gamma}}{2}
\right),
\qquad
\varepsilon^{\mu\nu\rho}
=\frac{\epsilon^{\mu\nu\rho}}{\sqrt{-g}},
\label{eq:euler-current}
\end{equation}
where $R_{\nu\rho\beta\gamma}$ is the Riemann tensor.
The term quadratic in \(\nabla_\mu u_\nu\) and the curvature contribution are not separately conserved, but for a unit timelike \(u^\mu\) their combination in Eq.~\eqref{eq:euler-current} is identically conserved off shell, \(\nabla_\mu J^\mu=0\). For a field of norm $\sigma=\pm1$ the curvature term carries the coefficient $+\tfrac{\sigma}{2}R_{\nu\rho\beta\gamma}$, with $[\nabla_\mu,\nabla_\nu]V^\rho=R^\rho{}_{\lambda\mu\nu}V^\lambda$; this is the origin of $\sigma$ in Eq.~\eqref{eq:euler-lagrangian}.

Consequently, one can couple \(J^\mu\) to a background electromagnetic
field without violating gauge invariance,
\begin{equation}
S_{\rm WZ}
=\kappa\int_{\mathcal M}d^3x\,\sqrt{-g}\;A_\mu J^\mu .
\label{eq:euler-coupling}
\end{equation}
The name ``Euler current'' follows from its integrated charge. If
\(u^\mu\) is the unit normal to a closed spacelike surface \(\Sigma\),
the Gauss-Codazzi relation gives
\begin{equation}
u_\mu J^\mu=\frac{-1}{8\pi}\,{}^{(2)}\!R,
\end{equation}
where ${}^{(2)}\!R$ is the intrinsic scalar curvature of $\Sigma$.
By the Gauss-Bonnet theorem,
\begin{equation}
\int_\Sigma d^2x\,\sqrt{h}\;u_\mu J^\mu
=\frac{-1}{2}\chi(\Sigma).
\label{eq:euler-charge}
\end{equation}

Let $n^\mu$ be the future-directed unit normal to the fixed surface $\Sigma$. For a smooth future-directed unit timelike field $u^\mu$, the conserved Euler-current charge is
\begin{equation}
 Q_J[\Sigma;u]=-\int_\Sigma d^2x\,\sqrt h\,n_\mu J^\mu[u]
 =\frac12\chi(\Sigma).
\end{equation}
It is unchanged under smooth deformations of $u^\mu$ within this time orientation; the local Euler-density relation above applies when $u=n$~\cite{GolkarRobertsSonEuler2015}. The induced electromagnetic charge is therefore
\begin{equation}
 Q_{\rm WZ}=\kappa Q_J=\frac{\kappa}{2}\chi(\Sigma).
 \label{eq:curvature-charge}
\end{equation}

For a relativistic quantum Hall state, the requisite timelike vector can
be constructed directly from a magnetically dominated electromagnetic
background, $u^\mu=\frac{1}{2b}\varepsilon^{\mu\nu\rho}F_{\nu\rho}$,
$b=\left(\frac12F_{\mu\nu}F^{\mu\nu}\right)^{1/2}$.
The Bianchi identity implies \(\nabla_\mu(bu^\mu)=0\), so that
\(bu^\mu\) is naturally interpreted as the magnetic-flux current. The
Euler-current term can then be regarded as a functional of \(A_\mu\) and
\(g_{\mu\nu}\) alone.
Including the electromagnetic Chern-Simons contribution, the total
charge on a closed surface is $Q=\nu N_\phi+\frac{\kappa}{2}\chi(\Sigma),
\qquad
N_\phi=\frac{1}{2\pi}\int_\Sigma F$.

The coefficient $\kappa$ is the curvature-dependent charge offset in the relativistic theory. For the Dirac zeroth-Landau-level states compared with non-relativistic states in Ref.~\cite{GolkarRobertsSon2014}, the sphere fluxes obey $N_\phi^{\rm NR}=N_\phi-1$. In that comparison,
$\kappa=\nu(\mathcal S_{\rm NR}-1)$, where $\mathcal S_{\rm NR}=2\bar s_{\rm NR}$ is the shift of the corresponding non-relativistic state.

Expanding the action around a constant magnetic background gives the Hall viscosity $\eta_{\rm H}=\frac{\kappa B}{8\pi}$~\cite{GolkarRobertsSon2014}.
The Euler-current coupling therefore unifies the curvature-induced charge, the relativistic shift, and the parity-odd viscosity response.
The Euler-current construction is diffeomorphism and local-Lorentz covariant, while a chosen fluid background has a preferred local rest frame. The unit timelike vector \(u^\mu\) is a covariant field, but any nonzero background value of \(u^\mu\) selects a timelike direction. This is physically appropriate for a quantum Hall fluid, whose finite density and magnetic background already break boost invariance.

\subsection{Extended Maxwell Algebra}

The Maxwell-algebra construction~\cite{Palumbo2016,DurkaKG2019} uses a Chern-Simons connection valued in an extension of the Poincar\'e algebra. Its independent fields are the coframe, Lorentz connection and an additional Lorentz-vector-valued one-form.

The Maxwell algebra in 2+1 dimensions is given by~\cite{BacryCombeRichard1970,Schrader1972,Concha2018,Salgado2019}
\begin{align}
 [J_a,J_b]&=\epsilon_{abc}J^c,
 & [J_a,P_b]&=\epsilon_{abc}P^c,
 & [J_a,Z_b]&=\epsilon_{abc}Z^c, \notag\\
 [P_a,P_b]&=\epsilon_{abc}Z^c,
 & [P_a,Z_b]&=0,
 & [Z_a,Z_b]&=0,
 \label{eq:maxwell}
\end{align}
with an invariant, non-degenerate
bilinear form satisfying
\begin{equation}
 \big\langle[X_A,X_B],X_C\big\rangle
 +\big\langle X_B,[X_A,X_C]\big\rangle=0 .
 \label{eq:invariance}
\end{equation}
For the Maxwell algebra, its non-vanishing components are given by
\begin{equation}
 \langle J_a,J_b\rangle=\alpha_0\eta_{ab},\qquad
 \langle J_a,P_b\rangle=\alpha_1\eta_{ab},\qquad
 \langle P_a,P_b\rangle
 =\langle J_a,Z_b\rangle=\alpha_2\eta_{ab},
 \label{eq:bilinear}
\end{equation}
with all other pairings vanishing. Nondegeneracy requires $\alpha_2\ne0$; coupling limits below are evaluated on the specified flat family.
Having specified the invariant bilinear form
\eqref{eq:bilinear}, we construct the Chern-Simons action associated with
the Maxwell algebra. For a Lie-algebra-valued connection \(\mathcal A\),
the action reads
\begin{equation}
 I_{\rm CS}[\mathcal A]
 =
 \frac{k}{4\pi}\int_{\mathcal M}
 \left\langle
 \mathcal A\wedge d\mathcal A
 +\frac{1}{3}\mathcal A\wedge[\mathcal A,\mathcal A]
 \right\rangle ,
 \label{eq:CSfunctional}
\end{equation}
where \(k\) is the Chern-Simons level and
\(\langle\cdot,\cdot\rangle\) denotes a nondegenerate invariant symmetric
bilinear form on the Maxwell algebra.
The gauge connection is decomposed according to the generators
\(\{J_a,P_a,Z_a\}\) as
\begin{equation}
 \mathcal A
 =
 \omega^a J_a+e^a P_a+A^a Z_a .
 \label{eq:connection}
\end{equation}
Here \(\omega^a\) is the dualized Lorentz spin connection, \(e^a\) is the
dreibein, and \(A^a\) is a Lorentz-vector-valued one-form associated with
the additional Maxwell generators \(Z_a\). The corresponding Lorentz
curvature and torsion two-forms are
\begin{equation}
 R^a=d\omega^a+\frac{1}{2}\epsilon^a{}_{bc}\,
 \omega^b\wedge\omega^c,
 \qquad
 T^a=de^a+\epsilon^a{}_{bc}\,
 \omega^b\wedge e^c .
 \label{eq:curvature-torsion}
\end{equation}

The bracket $[P_a,P_b]=\epsilon_{ab}{}^c Z_c$ relates internal gauge generators. It does not imply noncommuting spacetime coordinates or a noncommutative tangent space. An invertible coframe identifies tangent vectors with an internal Lorentz-vector space and defines $g_{\mu\nu}=\eta_{ab}e^a{}_\mu e^b{}_\nu$.

By employing the invariant tensor \eqref{eq:bilinear}, and discarding boundary
terms, the Chern-Simons functional becomes \cite{Palumbo2016,Concha2018,Salgado2019,Aviles2026}
\begin{equation}
 \begin{split}
 I=\frac{k}{4\pi}\int_{\mathcal M}\Big[&
 \alpha_0\Big(\omega^a\wedge d\omega_a
 +\frac13\epsilon_{abc}\omega^a\wedge\omega^b\wedge\omega^c\Big)
 +2\alpha_1e_a\wedge R^a\\
 &+\alpha_2\big(e_a\wedge T^a+2A_a\wedge R^a\big)\Big].
 \end{split}
 \label{eq:fullaction}
\end{equation}
The first term is the exotic Lorentz Chern-Simons action, the second is
the Einstein-Hilbert term in first-order form, and the
\(\alpha_2\)-sector contains both the torsional contribution
\(e_a\wedge T^a\) and the coupling between the Maxwell gauge field and
the Lorentz curvature.
In particular, the last contribution contains the topological term
\begin{equation}
 S_{\rm WZ}
 =
 \beta\int_{\mathcal M} A^a\wedge R_a,
 \qquad
 \beta=\frac{k\alpha_2}{2\pi}.
 \label{eq:WZderived}
\end{equation}
Equation~\eqref{eq:WZderived} is the Lorentz-covariant Wen-Zee-type coupling studied below. Its precise quantum-Hall interpretation requires an identification of the physical electromagnetic sector and of the geometric response coefficient in an appropriate magnetized background 
\cite{Palumbo2016,DurkaKG2019}.

This theory admits locally flat cosmological solutions carrying a cosmological horizon~\cite{Aviles2026}, which provide a setting in which the horizon contribution of the mixed gauge-curvature coupling can be
evaluated explicitly.
\subsection{AdS-Lorentz extension}  
\label{sec:adsl-algebra}
We also consider AdS-Lorentz Chern-Simons gravity, which admits black-hole solutions with the BTZ metric and nonvanishing torsion~\cite{AvilesHidalgoValdivia2023}. This provides a non-trivial setting to investigate the contribution of the mixed gauge-curvature coupling to black-hole entropy.

The Maxwell algebra admits a semisimple deformation known as the
AdS-Lorentz algebra ~\cite{Diaz2012,Concha2019,AvilesHidalgoValdivia2023}. Its nonvanishing
commutation relations are
\begin{equation}
\begin{aligned}
 \relax[J_a,J_b]&=\epsilon_{ab}{}^{c}J_c,
&
 [J_a,P_b]&=\epsilon_{ab}{}^{c}P_c,
&
 [J_a,Z_b]&=\epsilon_{ab}{}^{c}Z_c,
\\
 [P_a,P_b]&=\epsilon_{ab}{}^{c}Z_c,
&
 [Z_a,P_b]&=\epsilon_{ab}{}^{c}P_c,
&
 [Z_a,Z_b]&=\epsilon_{ab}{}^{c}Z_c .
\end{aligned}
\label{eq:adsl-algebra}
\end{equation}
The generators \(J_a\), \(P_a\), and \(Z_a\) correspond, respectively,
to Lorentz rotations, translations, and the additional AdS-Lorentz
sector. In contrast with the Maxwell algebra, the generators \(Z_a\) are
non-Abelian and act nontrivially on translations.
A symmetric invariant bilinear form is specified by
\begin{equation}
\begin{aligned}
 \langle J_a,J_b\rangle &=\alpha_0\eta_{ab},
&
 \langle J_a,P_b\rangle &=\alpha_1\eta_{ab},
&
 \langle J_a,Z_b\rangle &=\alpha_2\eta_{ab},
\\
 \langle P_a,P_b\rangle &=\alpha_2\eta_{ab},
&
 \langle P_a,Z_b\rangle &=\alpha_1\eta_{ab},
&
 \langle Z_a,Z_b\rangle &=\alpha_2\eta_{ab}.
\end{aligned}
\label{eq:AdSLorentzpairing}
\end{equation}
The existence of a nondegenerate invariant tensor reflects the
semisimple character of the AdS-Lorentz algebra and allows for a
Chern-Simons formulation. We assume $(\alpha_0-\alpha_2)(\alpha_2+\alpha_1)(\alpha_2-\alpha_1)\ne0$ when inferring flatness from the field equations. The explicit flat solutions can also be evaluated at degenerate coupling limits.
The AdS-Lorentz connection is taken to be
\begin{equation}
 \mathcal A
 =
 \omega^a J_a+\bar e^a P_a+A^a Z_a,
\label{eq:adsl-connection}
\end{equation}
where \(\omega^a\) is the dualized spin connection and \(A^a\) is the
one-form gauge field associated with the generators \(Z_a\).  We retain
the AdS radius \(\ell\) explicitly and define the rescaled dreibein by $\bar e^a\equiv\frac{e^a}{\ell}$,
where \(e^a\) is the physical dreibein.  The algebra and invariant
bilinear form above are unchanged; the length scale enters through
\(\bar e^a\).
The corresponding curvature two-form is
\begin{equation}
 \mathcal F
 =
 d\mathcal A+\frac{1}{2}[\mathcal A,\mathcal A]
 =
 R^a J_a+\mathcal T^a P_a+\mathcal F_Z^a Z_a ,
\label{eq:adsl-curvature}
\end{equation}
with
\begin{equation}
\begin{aligned}
 R^a &=
 d\omega^a+\frac{1}{2}\epsilon^a{}_{bc}
 \omega^b\wedge\omega^c,
\\
 \mathcal T^a &=
 \bar T^a+\epsilon^a{}_{bc}A^b\wedge \bar e^c,
\qquad
 \bar T^a\equiv\frac{T^a}{\ell}=
 d\bar e^a+\epsilon^a{}_{bc}\omega^b\wedge \bar e^c,
\\
 \mathcal F_Z^a &=
 D_\omega A^a
 +\frac{1}{2}\epsilon^a{}_{bc}\bar e^b\wedge \bar e^c
 +\frac{1}{2}\epsilon^a{}_{bc}A^b\wedge A^c,
\\
 D_\omega A^a &=
 dA^a+\epsilon^a{}_{bc}\omega^b\wedge A^c
\end{aligned}
\label{eq:adsl-field-strengths}
\end{equation}
Here $T^a=D_\omega e^a$ is the torsion of the physical dreibein. The $P$-sector curvature is $\mathcal T^a$. The field equations set $\mathcal T^a=0$, which does not make the geometry torsion-free but instead fixes $T^a=-\epsilon^a{}_{bc}A^b\wedge e^c$, so that $A^a$ sources spacetime
torsion whenever it is nonvanishing.

The additional terms in \(\mathcal T^a\) and \(\mathcal F_Z^a\) are a
direct consequence of the nonvanishing commutators
\([Z_a,P_b]\) and \([Z_a,Z_b]\).
The associated Chern-Simons action is
\begin{equation}
 I_{\rm AdSL}[\mathcal A]
 =
 \frac{k}{4\pi}\int_{\mathcal M}
 \left\langle
 \mathcal A\wedge d\mathcal A
 +\frac{1}{3}\mathcal A\wedge[\mathcal A,\mathcal A]
 \right\rangle .
\label{eq:adsl-CS-functional}
\end{equation}
Using the invariant tensor \eqref{eq:AdSLorentzpairing}, the action can be
written, up to an exact three-form, as \cite{AvilesHidalgoValdivia2023}
\begin{equation}
\begin{aligned}
 I_{\rm AdSL}
 =
 \frac{k}{4\pi}\int_{\mathcal M}
 \Bigg\{
 &
 \alpha_0\left(
 \omega^a\wedge d\omega_a
 +\frac{1}{3}\epsilon_{abc}
 \omega^a\wedge\omega^b\wedge\omega^c
 \right)
\\
 &+
 \alpha_1\left[
 2\bar e_a\wedge\left(R^a+D_\omega A^a\right)
 +\epsilon_{abc}\bar e^a\wedge A^b\wedge A^c
 +\frac{1}{3}\epsilon_{abc}\bar e^a\wedge \bar e^b\wedge \bar e^c
 \right]
\\
 &+
 \alpha_2\left[
 \bar e_a\wedge\bar T^a
 +2A_a\wedge R^a
 +A_a\wedge D_\omega A^a\right.
\\
 &\left.\qquad
 +\epsilon_{abc}A^a\wedge \bar e^b\wedge \bar e^c
 +\frac{1}{3}\epsilon_{abc}A^a\wedge A^b\wedge A^c
 \right]
 \Bigg\}
\end{aligned}
\label{eq:adsl-full-action}
\end{equation}

The \(\alpha_0\)-sector is the exotic Lorentz Chern-Simons term. The
\(\alpha_1\)-sector contains the Einstein-Hilbert term
\(2\bar e_a\wedge R^a\), together with a cosmological contribution and
non-minimal couplings to the AdS-Lorentz gauge field. The
\(\alpha_2\)-sector includes the torsional term \(\bar e_a\wedge \bar T^a\), the
curvature coupling \(2A_a\wedge R^a\), and the self-interactions required
by the non-Abelian \(Z_a\) sector.
In particular, one again identifies the term
\begin{equation}
 S_{\rm WZ}
 =
 \frac{k\alpha_2}{2\pi}\int_{\mathcal M} A_a\wedge R^a .
\label{eq:adsl-WZ-term}
\end{equation}

This is the direct AdS-Lorentz counterpart of the
Maxwell-algebra-induced relativistic Wen-Zee-type term. Unlike in the
Maxwell case, however, it is embedded in the larger gauge-invariant
combination displayed in \eqref{eq:adsl-full-action}; the terms
\(A_a\wedge D_\omega A^a\),
\(\epsilon_{abc}A^a\wedge \bar e^b\wedge \bar e^c\), and
\(\epsilon_{abc}A^a\wedge A^b\wedge A^c\) are required by the
AdS-Lorentz symmetry.

\vspace{0.3cm}

Thus, we have identified a relativistic version of the first Wen-Zee term that can be placed on any curved spacetime, including one carrying a Killing horizon. On the flat Maxwell and AdS-Lorentz connections studied here, $R^a=0$, so this particular mixed term vanishes pointwise on shell. This does not imply that the complete Chern-Simons Lagrangian vanishes. The Wald entropy, however, is built from $\partial\mathcal{L}/\partial R^{a}$ and can be evaluated before the field equations are imposed, and need not vanish with it. In the next section, we will compute the Wald entropy for this relativistic term on various backgrounds.

\section{Wald Entropy due to Wen-Zee mixed gauge-curvature coupling}
\label{sec:wald-wz}

For a stationary black hole with a bifurcate Killing horizon, the Wald
formalism identifies the entropy with the Noether charge associated with
the horizon-generating Killing vector field \(\xi^\mu\)
\cite{Wald1993,IyerWald1994}. In a generally covariant theory whose Lagrangian density depends algebraically on the Riemann tensor, the
entropy can be expressed as
\begin{equation}
 S_{\rm Wald}
 =
 -2\pi\oint_{\mathcal B}
 \frac{\partial\mathcal L}
 {\partial R_{\mu\nu\rho\sigma}}\,
 \varepsilon^\perp_{\mu\nu}
 \varepsilon^\perp_{\rho\sigma}\,
 \sqrt{h}\,d^{d-2}x ,
 \label{eq:wald-general}
\end{equation}
where \(\mathcal B\) is the bifurcation surface,
\(\varepsilon^\perp_{\mu\nu}\) is its normalized binormal, and \(h\) is
the determinant of the induced metric on \(\mathcal B\).
In \(2+1\) dimensions, the bifurcation surface of a stationary circular
black hole is a closed one-dimensional curve,
\(\mathcal B\simeq S^1\). Consequently, the Wald integral reduces to a
line integral,
\begin{equation}
 S_{\rm Wald}
 =
 -2\pi\oint_{\mathcal B}
 \frac{\partial\mathcal L}
 {\partial R_{\mu\nu\rho\sigma}}\,
 \varepsilon^\perp_{\mu\nu}
 \varepsilon^\perp_{\rho\sigma}\,
 \sqrt{g_{\phi\phi}}\,d\phi .
 \label{eq:wald-three-dimensional}
\end{equation}
Although the horizon ``area'' is therefore its circumference, the entropy
need not be proportional to this circumference. Any coupling that depends
explicitly on curvature may generate an additional contribution governed
by the pullback of the relevant background fields to \(\mathcal B\).
In the present case, the relativistic Wen-Zee term produces such a
correction as we show below.

\subsection{Contribution from the Wen-Zee term}
We calculate the horizon entropy associated to the relativistic Wen-Zee term induced from gauging the Maxwell algebra as discussed in the previous section.
We assume throughout that \(e^a{}_\mu\) is invertible,
so that \(g_{\mu\nu}=\eta_{ab}e^a{}_\mu e^b{}_\nu\) is a Lorentzian metric and the frame components \(\xi^a=\iota_\xi e^a\) determine
\(\xi^\mu\) uniquely.
Starting from the Lagrangian\begin{equation}
 L_{\rm WZ}=\beta A^a\wedge R_a,
\end{equation}
we can then find the pre-symplectic potential as discussed in detail in Appendix \ref{app:noether},
\begin{equation}
    \Theta(\delta)=-\beta A^a\wedge\delta\omega_a.
\end{equation}
We now follow the Lorentz-diffeomorphism construction of
Ref.~\cite{JacobsonMohd}. In a first-order formulation, the dynamical variables include the dreibein \(e^a=e^a{}_\mu dx^\mu\) and the dualized
Lorentz connection \(\omega^a=\frac12\epsilon^a{}_{bc}\omega^{bc}\).
A nondegenerate dreibein cannot be both regular at the bifurcation surface and invariant under the ordinary Lie derivative along the horizon Killing vector \(\xi=\xi^\mu\partial_\mu\). In
particular, although the Killing vector vanishes on the bifurcation surface, its derivative does not, and a regular frame is generally not preserved by the ordinary Lie derivative \(\mathcal L_\xi e^a\); choosing an invariant frame instead makes the connection singular there~\cite{JacobsonMohd}.

The appropriate symmetry is therefore a combination of a diffeomorphism
and a compensating local Lorentz transformation,
\begin{equation}
 \widehat\delta_{\xi,\rho}
 =
 \mathcal L_\xi+\delta_\rho .
 \label{eq:combined-lorentz-diffeomorphism}
\end{equation}
Here \(\mathcal L_\xi\) denotes the spacetime Lie derivative along the
horizon-generating Killing vector \(\xi^\mu\), while
\(\delta_\rho\) is an infinitesimal local Lorentz transformation with
parameter \(\rho^a\). In three dimensions, an antisymmetric Lorentz
parameter \(\rho^{ab}=-\rho^{ba}\) can be dualized into a Lorentz vector,
\(\rho^a=\tfrac12\epsilon^a{}_{bc}\rho^{bc}\).
With the convention adopted here, the Lorentz transformation acts on the
dreibein and spin connection as
\begin{equation}
 \delta_\rho e^a
 =
 -\epsilon^a{}_{bc}\rho^b e^c,
 \qquad
 \delta_\rho\omega^a
 =
 +D\rho^a,
 \qquad
 D\rho^a
 \equiv
 d\rho^a+\epsilon^a{}_{bc}\omega^b\rho^c ,
 \label{eq:lorentz-transformation-fields}
\end{equation}
with \(D\) the Lorentz-covariant exterior derivative. This is $\delta_\Lambda\mathcal A=\mathrm{d}\Lambda+[\mathcal A,\Lambda]$ with $\Lambda=\rho^aJ_a$; this convention is used throughout and fixes all signs below. For the AdS-Lorentz black hole, Sec.~\ref{sec:adsl-smooth} constructs a representative in which this Lorentz boost preserves the full connection near the bifurcation circle.

For $L_{\rm WZ}$ the Lorentz-diffeomorphism Noether charge is the one-form
\begin{equation}
 Q_{(\xi,\rho)}
 =
 \beta\left(\iota_\xi\omega^a+\rho^a\right)A_a.
 \label{eq:appC:combined}
\end{equation}
Here \(\beta\) is the Wen-Zee-type response coefficient,
\(A^a=A^a{}_\mu dx^\mu\) is the Lorentz-vector-valued gauge field, \(A_a=\eta_{ab}A^b\), and \(\iota_\xi\) denotes interior contraction with the vector field \(\xi\), so that
\(\iota_\xi\omega^a=\xi^\mu\omega^a{}_\mu\).
Since \(Q_{(\xi,\rho)}\) is a one-form in \(2+1\) dimensions, it can be
integrated over the one-dimensional bifurcation surface
\(\mathcal B\simeq S^1\).
Let \(\xi^\mu\) be normalized such that its surface gravity is \(\kappa_{\rm sg}\), namely
\(\xi^\nu\nabla_\nu\xi^\mu=\kappa_{\rm sg}\xi^\mu\) on the Killing horizon.

The Wald entropy contribution associated to the Wen-Zee-type sector is then
\begin{equation}
 S_{\rm WZ}
 =
 \frac{2\pi}{\kappa_{\rm sg}}
 \oint_{\mathcal B}Q_{(\xi,\rho)}.
 \label{eq:WZ-noether-entropy}
\end{equation}
The integral must be evaluated on the bifurcation circle
\(\mathcal B\), rather than on a spatial hypersurface \(\Sigma\). At \(\mathcal B\) the horizon Killing field vanishes,
\(\xi^\mu|_{\mathcal B}=0\), and assuming both the frame and the independent connection are regular there, one has \(\iota_\xi\omega^a|_{\mathcal B}=0\).
Only the compensating Lorentz transformation contributes to the horizon
charge:
\begin{equation}
 S_{\rm WZ}
 =
 \frac{2\pi\beta}{\kappa_{\rm sg}}
 \oint_{\mathcal B}\rho^aA_a.
 \label{eq:WZ-compensator-form}
\end{equation}

The compensator \(\rho^a\) is fixed by requiring the dreibein to be invariant under the combined transformation at the bifurcation surface, \(\widehat\delta_{\xi,\rho}e^a|_{\mathcal B}=0\)  as shown in Appendix~\ref{sec:fixinglorentzcompensatormaxwell}, which gives 
\begin{equation}
 \rho^a
 =
 -\kappa_{\rm sg}\widetilde n^a,
 \qquad
 \widetilde n^a
 \equiv
 \frac12\epsilon^{abc}n_{bc}.
 \label{eq:compensator}
\end{equation}
Here \(n_{ab}\) is the internal binormal to the bifurcation surface and
\(\widetilde n^a\) is its dual Lorentz vector.

In three dimensions, the dual of the binormal is tangent to the one-dimensional bifurcation surface. For a circular horizon parametrized by \(\phi\), with induced line element
\begin{equation}
 ds^2_{\mathcal B}
 =
 g_{\phi\phi}\,d\phi^2,
 \label{eq:bifurcation-induced-metric}
\end{equation}
the normalized internal tangent can be written as
\begin{equation}
 \widetilde n^a
 =
 \frac{- \; e^a{}_\phi}{\sqrt{g_{\phi\phi}}}.
 \label{eq:dual-binormal-tangent}
\end{equation}
We take $\epsilon_{012}=+1$, orient the binormal by $\nabla^a\xi^b|_{\mathcal B}=\kappa_{\rm sg}n^{ab}$, and choose the angular orientation so that its dual points opposite to $+\partial_\phi$. The explicit horizon frames below realize this convention.

Substitution of \eqref{eq:compensator} into
\eqref{eq:WZ-compensator-form} yields
\begin{equation}
 S_{\rm WZ}
 =
 -2\pi\beta\oint_{\mathcal B}\widetilde n^aA_a.
 \label{eq:WZ-projected}
\end{equation}

Equivalently, in the angular parametrization of \(\mathcal B\),
\begin{equation}
 S_{\rm WZ}
 =
 2\pi\beta\oint_{\mathcal B}
 \frac{e^a{}_\phi A_{a\phi}}
 {\sqrt{g_{\phi\phi}}}\,d\phi,
 \qquad
 A_{a\phi}\equiv A_{a\mu}
 \frac{\partial x^\mu}{\partial\phi}.
 \label{eq:WZ-projected-components}
\end{equation}

This is the central result of the Lorentz-compensator analysis: for smooth fields admitting the specified Lorentz symmetry, the mixed coupling contributes the line integral of $A^a$ projected along the dual binormal selected by the horizon geometry.

For a Lorentz-valued lift the frame condition fixes the boost parameter even with smooth torsion, but the same combined transformation must preserve the other dynamical fields. When other generators also rotate the frame, as $Z_a$ does in the AdS-Lorentz algebra, the frame fixes only the total rotation, and the Lorentz-valued lift is a choice (Sec.~\ref{sec:adsl-lift}). This is checked explicitly for the backgrounds below. When $A^a$ is an external source, the expression is its contribution to the local horizon-charge functional; a thermodynamic first law additionally requires the full solution and a compatible prescription for varying the source.

\subsection{Holonomy interpretation of Wen-Zee contribution to Wald entropy}
\label{sec:holonomy}

The contribution $\beta A_a$ is the momentum conjugate to Lorentz curvature. At a regular bifurcation circle the horizon symmetry is a normal-plane boost. Equation~\eqref{eq:WZ-projected} is $2\pi$ times its unit-boost surface charge. Therefore the Wen-Zee coupling endows the horizon-generating Lorentz boost with a surface charge. The boost acts in the plane normal to the bifurcation circle, and its dual $\widetilde n^a$ selects the component of the gauge field that contributes to this charge.

This also has a response-theory counterpart in the non-relativistic Wen-Zee term. Variation of $A_{EM}\wedge\mathrm{d}\omega$ with respect to
the electromagnetic potential gives curvature-bound electric charge.
Differentiation of the first-order gravitational coupling with respect
to Lorentz curvature instead gives the boost surface charge.
Both are mixed gauge-curvature responses in different settings and thus have charges with different physical identity.

Further defining $a_{\mathcal B}=\widetilde n_aA^a|_{\mathcal B}$, the
entropy contribution is written as 
\begin{equation}
 \Delta S_{\rm WZ}
 =-2\pi\beta\oint_{\mathcal B}a_{\mathcal B}.
\end{equation}

On a smooth bifurcation circle, $\mathring D_{\mathcal B}\widetilde n=0$,
because $\mathcal B$, the fixed-point set of the flow of $\xi$, is a
closed geodesic and $\widetilde n$ is its unit tangent.
Consequently $D_{\mathcal B}\widetilde n=0$ whenever the pulled-back
contorsion is aligned with $\widetilde n$, including the torsion-free case
and the regular AdS-Lorentz representative constructed below.
For the Maxwell algebra, this condition ensures that a $Z$ shift induces $a_{\mathcal B}\mapsto a_{\mathcal B}
+d_{\mathcal B}(\widetilde n\cdot\zeta)$ (Appendix~\ref{app:abelian-connection}).
Its circulation is therefore invariant under single-valued shifts
and measures an additive Abelian holonomy around the horizon. This gives the result a direct connection to Aharonov-Bohm
physics: given a compatible global gauge group and a probe charge representation $q$, the same circulation enters the Wilson factor $\exp(iq\oint_{\mathcal B}a_{\mathcal B})$.
The entropy measures this circulation as a real boost charge; an interference experiment would instead measure its exponentiated
phase. Establishing a force-free Aharonov-Bohm experiment requires a suitable probe trajectory and field configuration, but the
underlying connection to gauge circulation is already explicit in the horizon charge.

\subsection{Curvature response from Euler current-curvature coupling}
\label{sec:eulerentropy}

We examine the explicit contribution of the Euler current-curvature coupling term to the Wald functional. 
Only the explicitly curvature-dependent term in
\eqref{eq:euler-current} survives the curvature variation entering \eqref{eq:wald-general}, namely
\begin{equation}
 \mathcal L_{\rm EC}
 =\frac{\kappa\sigma}{16\pi}\,A_\mu\,\varepsilon^{\mu\nu\rho}\varepsilon^{\alpha\beta\gamma}
 u_\alpha R_{\nu\rho\beta\gamma},
 \label{eq:euler-lagrangian}
\end{equation}
The component evaluation, including conventions, is given in
Appendix~\ref{app:euler-current-derivation}.  
Writing $\hat t^\mu=\delta^\mu_\phi/\sqrt{g_{\phi\phi}}$ and
$\mathrm{d} s=\sqrt{g_{\phi\phi}}\,\mathrm{d}\phi$, we obtain
\begin{equation}
 S_{\rm EC}
 =-\frac{\kappa\sigma}{2}\oint_{\mathcal{B}}
 \big(A\cdot\hat t\big)\big(u\cdot\hat t\big)\,\mathrm{d} s
 \;=\;
 +\frac{\kappa}{2}\oint_{\mathcal{B}}
 \big(A\cdot\hat t\big)\big(u\cdot\hat t\big)\,\mathrm{d} s,
 \label{eq:eulerentropyinvariantform}
\end{equation}
with $\sigma=-1$. 
This calculation identifies its local curvature response but its interpretation as equilibrium horizon entropy requires an additional regularity condition.

To compare the local curvature responses of the two constructions, we restrict the
Lorentz-vector-valued gauge field \(A^a\) to the effective configuration
\begin{equation}
 A^a_{\rm eff}=A\,u^a,
 \label{eq:euler-effective-probe}
\end{equation}
where \(A=A_\mu dx^\mu\) is the electromagnetic \(U(1)\) gauge-field
one-form and
\begin{equation}
 u^a=e^a{}_\mu u^\mu,
 \qquad
 u^au_a=-1,
 \label{eq:frame-fluid-velocity}
\end{equation}
are the frame components of the unit fluid velocity. Equation
\eqref{eq:euler-effective-probe} should be understood as an effective
embedding of the electromagnetic background into the Lorentz-vector-valued
gauge sector. It is imposed only for the purpose of comparing the
Wen-Zee-type horizon responses and does not assert a general off-shell
identification of the two bulk theories.

Substituting \eqref{eq:euler-effective-probe} into the first-order entropy
formula \eqref{eq:WZ-projected}, and restricting to the bifurcation circle
\(\mathcal B\), gives
\begin{equation}
\begin{aligned}
 S_{\rm WZ}
 &=
 -2\pi\beta\oint_{\mathcal B}
 \widetilde n_a A^a_{\rm eff}
 \\
 &=
 -2\pi\beta\oint_{\mathcal B}
 (\widetilde n\cdot u)\,A_\phi\,d\phi .
\end{aligned}
\label{eq:WZ-effective-entropy}
\end{equation}
Here \(A_\phi=A_\mu(\partial x^\mu/\partial\phi)\) is the pullback of the
electromagnetic gauge field to \(\mathcal B\) and
\(\widetilde n^a\) is the internal unit tangent to the bifurcation circle.
With \(\widetilde n^a=-e^a{}_\phi/\sqrt{g_{\phi\phi}}\) and
\(\hat t^\mu=\delta^\mu{}_\phi/\sqrt{g_{\phi\phi}}\),
one finds
\begin{equation}
 \widetilde n\cdot u
 =
 \eta_{ab}\widetilde n^au^b
 =
 -\frac{e^a{}_\phi u_a}{\sqrt{g_{\phi\phi}}}
 =
 -\frac{u_\phi}{\sqrt{g_{\phi\phi}}}
 =
- u\cdot\hat t.
 \label{eq:internal-metric-contraction}
\end{equation}
Since \(ds=\sqrt{g_{\phi\phi}}\,d\phi \),
\begin{equation}
 A_\phi\,d\phi
 =
 \left(\frac{A_\phi}{\sqrt{g_{\phi\phi}}}\right)ds
 =
 (A\cdot\hat t)\,ds.
 \label{eq:gauge-pullback-invariant}
\end{equation}
The first-order entropy contribution can consequently be expressed as
\begin{equation}
 S_{\rm WZ}
 =
 2\pi\beta\oint_{\mathcal B}
 (u\cdot\hat t)(A\cdot\hat t)\,ds.
 \label{eq:WZ-entropy-invariant}
\end{equation}

For \(\sigma=-1\), the Euler-current calculation gave
\begin{equation}
 S_{\rm EC}
 =
 \frac{\kappa}{2}\oint_{\mathcal B}
 (u\cdot\hat t)(A\cdot\hat t)\,ds.
 \label{eq:eulerentropyinvariantform-bis}
\end{equation}
The two expressions have the same functional form and coincide provided
$2\pi\beta=\kappa/2$, i.e.\ $\beta=\kappa/4\pi$.
This comparison concerns the explicit curvature responses on Eq.~\eqref{eq:euler-effective-probe}. Under an electromagnetic gauge transformation $A\mapsto A+\mathrm{d}\lambda$, the formal Euler-current expression varies by $-\frac{\kappa}{2}\oint_{\mathcal B}\lambda\,\mathrm{d}_{\mathcal B}(u\cdot\hat t)$; a complete horizon charge requires the full theory and its regularity conditions.

The Euler-current formulation nevertheless carries a structural obstruction at a bifurcate horizon. Smoothness and stationarity of $u^\mu$ imply $0=(\mathcal L_\xi u)^\mu\big|_{\mathcal B}
 =-\kappa_{\rm sg}u^\nu
 \varepsilon^\perp_\nu{}^\mu\big|_{\mathcal B}$,
where $\kappa_{\rm sg}\neq0$ is the surface gravity.
Thus $u^\mu$ would have to be tangent to the spacelike circle, in contradiction with $u_\mu u^\mu=-1$.
This is the regularity obstruction familiar from stationary unit timelike fields.

We therefore retain Eq.~\eqref{eq:eulerentropyinvariantform}
as a formal local curvature contribution, rather than an established equilibrium entropy of the full Euler-current theory. A construction using fields regular only on the future horizon requires a separate analysis of the complete horizon surface-charge variation. The first-order Maxwell coupling does not require such a timelike vector field.
In the quantum Hall realization, moreover, $u^\mu=\varepsilon^{\mu\nu\rho}F_{\nu\rho}/2b$ with
$b=(\tfrac12F_{\mu\nu}F^{\mu\nu})^{1/2}$ exists only where the
magnetic-domination condition $\tfrac12F_{\mu\nu}F^{\mu\nu}>0$ holds.

\section{Wen-Zee contribution to the horizon entropy in Maxwell and AdS-Lorentz Chern-Simons gravity}
\label{sec:entropy:maxwell-comparison}

Before proceeding, we note that the physical interpretations of the Maxwell gauge fields in the quantum Hall systems and Maxwell Chern-Simons gravity theories are different. In the relativistic quantum
Hall effective description, the Lorentz-vector-valued field is restricted to an effective configuration constructed from the electromagnetic gauge field. In Maxwell Chern-Simons gravity, instead, the corresponding field is the dynamical gauge field associated with the Maxwell generators of the gravitational connection.

The comparison therefore identifies $A^a$ field with the gravitational Maxwell field on the relevant classical background. 

The resulting entropy is controlled by its component projected along the dual binormal, which is tangent to the bifurcation circle. We evaluate this projection on the following classical backgrounds.

\subsection{The flat cosmological spacetime}
\label{sec:entropy:background}

For later use, we introduce two equivalent conventions for the tangent-space
metric,
\begin{equation}
 \text{(i)}\qquad
 \eta_{ab}=\operatorname{diag}(-1,1,1),
 \qquad
 \text{(ii)}\qquad
 \eta_{01}=\eta_{10}=\eta_{22}=1,
 \label{eq:etaconventions}
\end{equation}
with all remaining components of \(\eta_{ab}\) vanishing in convention
(ii). The first choice is the standard orthonormal Lorentz basis. The
second is a null basis, for which the scalar product of two Lorentz
vectors takes the form $V_aU^a =V^0U^1+V^1U^0+V^2U^2$.
All covariant expressions derived below are independent of this choice.
Nevertheless, we employ the null basis in the explicit evaluation in
order to facilitate direct comparison with Ref.~\cite{Aviles2026}.
We consider the locally flat cosmological solution characterized by the constant parameters \(\{\mathcal M,\mathcal N,\mathcal F\}\), as studied in Ref.~\cite{Aviles2026}. In the Chern-Simons formulation, the solution is specified by the angular component of the reduced connection,
\begin{equation}
 a_\phi
 =
 J_1+\frac{\mathcal M}{2}J_0
 +\frac{\mathcal N}{2}P_0
 +\frac{\mathcal F}{2}Z_0.
 \label{eq:FSCconnection}
\end{equation}
The full radial dependence is recovered through the standard gauge
dressing of the reduced connection. The associated metric is locally flat
and possesses a cosmological horizon at
\begin{equation}
 r_c
 =
 \frac{|\mathcal N|}{2\sqrt{\mathcal M}},
 \qquad
 \mathcal M>0.
 \label{eq:rc}
\end{equation}
We restrict to $\mathcal M>0$ and $\mathcal N\neq0$, so that $r_c>0$ and the
circular horizon and its unit tangent are nondegenerate.

In Maxwell Chern-Simons gravity, the Lorentz-vector-valued field coupled
to the curvature is not an external probe, but the dynamical Maxwell
gauge field \(A^a\), namely the component of the full connection
associated with the generators \(Z_a\). For a nondegenerate invariant form, the full coupled field equations imply $R=T=0$ and the Maxwell-sector flatness condition
\begin{equation}
 DA^a
 +\frac12\epsilon^a{}_{bc}e^b\wedge e^c
 =
 0.
 \label{eq:gravitational-maxwell-equation}
\end{equation}
Accordingly, in order to compare the Wen-Zee-type horizon contribution
with the Maxwell Chern-Simons result, we specialize the first-order field
according to $A^a\longrightarrow \left.A^a\right|_{\rm on\ shell}$.
This identification concerns the first-order curvature coupling and
should not be confused with the effective electromagnetic embedding
\(A^a_{\rm eff}=A\,u^a\) employed previously to match the Euler-current
entropy.

In the null basis \eqref{eq:etaconventions}, the pullbacks of the relevant gauge
fields to the cosmological-horizon circle, obtained by evaluating the
gauge-dressed connection of Ref.~\cite{Aviles2026} at $r=r_c$, are
\begin{align}
 \omega^a{}_\phi\big|_{r_c}
 &=
 \left(
 \frac{\mathcal M}{2},\,1,\,0
 \right),
 \notag\\[2pt]
 e^a{}_\phi\big|_{r_c}
 &=
 \left(
 \frac{\mathcal N}{4},\,
 \frac{\mathcal N}{2\mathcal M},\,0
 \right),
 \notag\\[2pt]
 A^a{}_\phi\big|_{r_c}
 &=
 \left(
 \frac{\Xi}{4},\,
 \frac{\Xi}{2\mathcal M},\,0
 \right),
 \qquad
 \Xi
 \equiv
 \mathcal F-\frac{\mathcal N^2}{4\mathcal M}.
 \label{eq:horizondata}
\end{align}
Appendix~\ref{app:fsc-evaluation} constructs fields smooth at the bifurcation circle whose angular pullbacks are related to these by a constant Lorentz rotation; this justifies evaluating the Wald charge at $\mathcal B$ with the stationary data of Ref.~\cite{Aviles2026}. The horizon dreibein satisfies $e_{a\phi}e^a{}_\phi=r_c^2$,
which agrees with the induced metric on the cosmological-horizon circle.
The entropy contribution of the Maxwell sector is determined by the
projection of \(A^a\) along the dual binormal
\(\widetilde n^a\). For the solution \eqref{eq:horizondata}, one finds
\begin{equation}
 \widetilde n^aA_{a\phi}
 =-
 \operatorname{sgn}(\mathcal N)\,
 \frac12\left(
 \frac{\mathcal F}{\sqrt{\mathcal M}}
 -\frac{\mathcal N^2}{4\mathcal M^{3/2}}
 \right).
 \label{eq:FSCprojection}
\end{equation}
The details of this component calculation are reported in
Appendix~\ref{app:fsc-evaluation}. Substitution into
Eq.~\eqref{eq:WZ-projected}, together with
\(\beta=k\alpha_2/(2\pi)\), yields
\begin{equation}
 S_{\alpha_2}
 =
 \pi k\alpha_2\,
 \operatorname{sgn}(\mathcal N)
 \left(
 \frac{\mathcal F}{\sqrt{\mathcal M}}
 -\frac{\mathcal N^2}{4\mathcal M^{3/2}}
 \right).
 \label{eq:FSCresult}
\end{equation}
This is the $\alpha_2$ sector of the entropy obtained in the canonical approach in Ref.~\cite{Aviles2026}.

The Einstein-Hilbert sector provides an independent normalization check.
Replacing the Maxwell field by the dreibein in the projected charge,
together with \(\alpha_2\rightarrow\alpha_1\), gives
\begin{equation}
 S_{\rm EH}
 =
 2\pi k\alpha_1r_c
 =
 \pi k\alpha_1\frac{|\mathcal N|}{\sqrt{\mathcal M}}.
 \label{eq:FSCeinsteincheck}
\end{equation}
This reproduces the \(\alpha_1\)-dependent contribution to the canonical
entropy of the flat cosmological solution. Including the exotic CS sector,
with the same geometric boost and vanishing integration constant, gives
\begin{equation}
 S_{\rm FSC}=2\pi k\operatorname{sgn}(\mathcal N)
 \left[\alpha_0\sqrt{\mathcal M}
 +\frac{\alpha_1\mathcal N+\alpha_2\Xi}{2\sqrt{\mathcal M}}\right].
\end{equation}

\subsection{The AdS-Lorentz black hole in a regular horizon representative}
\label{sec:adsl-smooth}

For the AdS-Lorentz black hole of Ref.~\cite{AvilesHidalgoValdivia2023}, we construct fields that are smooth at the outer bifurcation circle and evaluate the charge there. A gauge representative is a connection chosen within the exterior Chern-Simons gauge orbit. The representative used below has the same BTZ metric and asymptotic fields as the original solution. Only its independent connection and contorsion differ. Appendix~\ref{app:regular-equivalence} gives the transformation to a gauge equivalent representative, and Appendix~\ref{app:bms-route} shows the obstruction in the original BMS fields.

\subsubsection{Geometry and the regular connection}
Define
\begin{equation}
 x=\frac{r_+}{\ell},\qquad y=\frac{r_-}{\ell},\qquad
 q=\sqrt{-b}>0,\qquad x>y\geq0,
 \label{eq:adsl-xy-conventions}
\end{equation}
where $b$ labels the independent flat-connection sector. The metric is
\begin{equation}
 \mathrm{d} s^2=-N^2\mathrm{d} t^2+\frac{\mathrm{d} r^2}{N^2}
 +r^2(\mathrm{d}\varphi+N^\varphi\mathrm{d} t)^2,
 \quad N^2=\frac{(r^2-r_+^2)(r^2-r_-^2)}{\ell^2r^2},
 \quad N^\varphi=-\frac{r_+r_-}{\ell r^2}.
 \label{eq:adsl-btz-metric}
\end{equation}
The horizon generator and its positive surface gravity are
\begin{equation}
 \xi=\partial_t+\Omega_H\partial_\varphi,\qquad
 \Omega_H=\frac{y}{\ell x},\qquad
 \kappa_{\rm sg}=\frac{x^2-y^2}{\ell x}.
 \label{eq:adsl-horizon-generator}
\end{equation}
The extremal limit and the parabolic sector $q=0$ are excluded from
this construction.

Introduce $r_*$ by $\mathrm{d} r_*/\mathrm{d} r=N^{-2}$ and use
\begin{equation}
 U=-e^{-\kappa_{\rm sg}(t-r_*)},\qquad
 V=e^{\kappa_{\rm sg}(t+r_*)},\qquad
 \widetilde\phi=\varphi-\Omega_Ht.
 \label{eq:adsl-kruskal-coordinates}
\end{equation}
The outer bifurcation circle is $\mathcal B=\{U=V=0\}$, and
$\xi=\kappa_{\rm sg}(V\partial_V-U\partial_U)$.
In the null internal basis a regular dreibein is
\begin{equation}
 \begin{split}
 e^0&=-\sqrt{\mathsf A/2}\,\mathrm{d} V,\qquad
 e^1=\sqrt{\mathsf A/2}\,\mathrm{d} U,\\
 e^2&=r\big[\mathrm{d}\widetilde\phi+
 \mathsf H(U\mathrm{d} V-V\mathrm{d} U)\big],
 \end{split}
 \label{eq:adsl-regular-frame}
\end{equation}
where $\mathsf A$ and $\mathsf H$ are smooth functions of $UV$, with
$\mathsf A(0)>0$. Their explicit expressions and the regularity proof
are given in Appendix~\ref{app:kruskal}. Let $\mathring{\omega}$ be the Levi-Civita
connection of this frame.

The AdS-Lorentz algebra decomposes into three commuting
$\mathfrak{sl}(2,\mathbb R)$ factors. Hence the generators and invariant
pairings diagonalize as
\begin{equation}
 \widehat J_a=J_a-Z_a,\qquad J_a^\pm=\tfrac12(Z_a\pm P_a),\qquad
 c_0=\alpha_0-\alpha_2,\qquad c_\pm=\tfrac12(\alpha_2\pm\alpha_1),
 \label{eq:adsl-factor-generators}
\end{equation}
with $\langle\widehat J_a,\widehat J_b\rangle=c_0\eta_{ab}$,
$\langle J_a^\pm,J_b^\pm\rangle=c_\pm\eta_{ab}$ and vanishing mixed
pairings. Thus
\begin{equation}
 \mathcal A=\omega^a\widehat J_a+
 \mathcal A_+^aJ_a^++\mathcal A_-^aJ_a^- ,\qquad
 \mathcal A_\pm=\mathring{\omega}\pm\bar e,\quad \bar e=e/\ell,\quad A=\mathring{\omega}-\omega.
 \label{eq:adsl-factor-connections}
\end{equation}
The $\pm$ connections are flat because the metric is locally AdS and
$D_{\mathring{\omega}}e=0$. The independent connection $\omega$ is flat separately and
is not fixed by the BTZ metric. For the independent flat connection
$\omega$, we choose the following horizon-adapted representative, whose
motivation is discussed in Appendix~\ref{app:regular-equivalence}:
\begin{equation}
 \omega=-s_0q\,\widetilde n\,\mathrm{d}\widetilde\phi,
 \qquad \widetilde n=(0,0,-1),\qquad s_0=\pm1.
 \label{eq:adsl-regular-independent}
\end{equation}
Here $\widetilde n$ is a constant internal vector extending the dual
binormal away from $\mathcal B$. The magnitude $q$ is fixed by the
invariant of the independent connection sector, while the sign $s_0$
specifies the relative orientation of the connection holonomy and the
horizon boost. Appendix~\ref{app:regular-equivalence} shows explicitly
that this representative is related by a gauge transformation to the
exterior connection of Ref.~\cite{AvilesHidalgoValdivia2023} and can be
matched to the original asymptotic fields.

\subsubsection{The Lift of the Horizon Symmetry}
\label{sec:adsl-lift}

A lift of the Killing symmetry combines the diffeomorphism generated by $\xi$ with a gauge transformation of parameter $\Lambda$, called the compensator. 
For a combined diffeomorphism and gauge transformation, define
\begin{equation}
 \Psi=\Lambda+\iota_\xi\mathcal A,\qquad
 (\mathcal{L}_\xi+\delta_\Lambda)\mathcal A
 =\iota_\xi\mathcal F+D_{\mathcal A}\Psi.
 \label{eq:bulk-exact-symmetry}
\end{equation}

An exact symmetry of a flat solution obeys $D_{\mathcal A}\Psi=0$. For smooth fields, $\Psi|_{\mathcal B}=\Lambda|_{\mathcal B}$, and the tangential condition is
\begin{equation}
 D_{\widetilde\phi}\Lambda|_{\mathcal B}=0.
 \label{eq:master-exact-symmetry}
\end{equation}

The frame~\eqref{eq:adsl-regular-frame} satisfies
$\mathcal{L}_\xi e=\rho\times e$, with the constant parameter
\begin{equation}
 \rho^a=-\kappa_{\rm sg}\widetilde n^a=(0,0,\kappa_{\rm sg}),\qquad
 \Lambda=\rho^aJ_a.
 \label{eq:adsl-geometric-lift}
\end{equation}
The Levi-Civita connection obeys $\mathcal{L}_\xi\mathring{\omega}=-\mathring D\rho$. The
form of the independent connection in~\eqref{eq:adsl-regular-independent}
gives two further properties. Since $\xi(\widetilde\phi)=0$,
$\mathcal{L}_\xi\omega=0$, while the alignment of $\omega$ with $\rho$ implies
$D_\omega\rho=0$. Consequently,~\eqref{eq:adsl-geometric-lift} is an
exact symmetry of the complete connection near the bifurcation circle.
The regularity of the fields and the alignment of the pulled-back
contorsion with $\widetilde n^a$ ensure that torsion does not obstruct
this lift, in contrast to the BMS-gauge representative discussed in
Appendix~\ref{app:bms-route}.

Since $\Lambda=\rho^aJ_a$ has no $P$ or $Z$ component and $J_a=\widehat J_a+J_a^++J_a^-$, this gives
\begin{equation}
 \Psi_0|_{\mathcal B}
 =\Psi_+|_{\mathcal B}
 =\Psi_-|_{\mathcal B}
 =-\kappa_{\rm sg}\widetilde n,
 \qquad
 \Psi_i\cdot\Psi_i=\kappa_{\rm sg}^2.
 \label{eq:adsl-thermal-norms}
\end{equation}
For each flat factor, exact symmetry gives $D\Psi_i=0$, and thus the
normalization fixed on $\mathcal B$ holds throughout the connected flat
region.

It is useful to express this normalization in terms of the
dimensionless thermal generator $h_i=2\pi\Psi_i/\kappa_{\rm sg}$. Using
$\operatorname{tr}(J_aJ_b)=\eta_{ab}/2$, we obtain
\begin{equation}
 \operatorname{tr}(h_i^2)=2\pi^2.
 \label{eq:euclidean-holonomy}
\end{equation}

Smoothness alone does not fix the independent-sector symmetry parameter. In the regular representative, $\Psi_0=C\widetilde n$ is covariantly constant for arbitrary $C$, while the $\pm$ parameters remain fixed. The resulting family of entropy expressions is
\begin{equation}
 S(C)=2\pi k\left[\alpha_1x-\alpha_2y
 -s_0\frac{C}{\kappa_{\rm sg}}(\alpha_0-\alpha_2)q\right].
 \label{eq:entropy-lift-family}
\end{equation}
The first law is satisfied for every $C$, with the corresponding independent-sector potential. In particular, $C=0$ gives the $Z$-valued lift $-\kappa_{\rm sg}\widetilde n^aZ_a$, which still rotates the frame because $[Z_a,P_b]=\epsilon_{ab}{}^cP_c$, but produces no $\sqrt{-b}$ contribution to the entropy.

We instead require the horizon boost to be generated by $J_a$ in all three factors. This fixes $C=-\kappa_{\rm sg}$ and gives the unit thermal normalization. This is an input: it identifies $J_a$, rather than $Z_a$, as the tangent-space Lorentz generator. The remaining sign $s_0=\pm1$ is not fixed by the metric or by regularity. We choose $s_0=+1$ as discrete ensemble data (Appendix~\ref{app:s0-branch}). Thus the entropy prescription requires, in addition to the connection, a specification of the thermal lift and of the discrete branch.

\subsubsection{Entropy at the bifurcation circle}
\label{sec:adsl-entropy}
At $\mathcal B$ the fields are regular and
$\iota_\xi\mathcal A|_{\mathcal B}=0$, so $\Psi|_{\mathcal B}=\Lambda|_{\mathcal B}$
and, as for the mixed term in Sec.~\ref{sec:wald-wz}, only the
compensator contributes to the horizon charge of the combined
transformation. The symmetry lift follows the first-order horizon construction of Refs.~\cite{JacobsonMohd,Prabhu2017}.

The horizon pullbacks in the regular frame, evaluated in
Appendix~\ref{app:regular-pullbacks}, are particularly simple:
\begin{equation}
 \bar e^a{}_\phi=-x\widetilde n^a,\qquad
 \mathring{\omega}^a{}_\phi=y\widetilde n^a,\qquad
 \omega^a{}_\phi=-s_0q\widetilde n^a,\qquad
 A^a{}_\phi=(y+s_0q)\widetilde n^a.
 \label{eq:adsl-smooth-pullbacks}
\end{equation}
Here and below $\phi$ on $\mathcal B$ denotes the regular angle
$\widetilde\phi$. Pairing the lift~\eqref{eq:adsl-geometric-lift} with
the connection through the invariant form~\eqref{eq:AdSLorentzpairing}
gives the charge
\begin{equation}
 Q_{\mathcal B}=\frac{k}{2\pi}\rho_a
       \big(\alpha_1\bar e^a+\alpha_0\omega^a+\alpha_2A^a\big),
 \qquad S=\frac{2\pi}{\kappa_{\rm sg}}\oint_{\mathcal B}Q_{\mathcal B}.
 \label{eq:sectorwise-total-charge}
\end{equation}
Its $\alpha_2$ part is the charge~\eqref{eq:WZ-compensator-form} of the
mixed coupling and its $\alpha_1$ part is the Einstein-Hilbert charge.
Using $\widetilde n^2=1$ gives, for $s_0=+1$,
\begin{equation}
 S=2\pi k\left[\alpha_1\frac{r_+}{\ell}
 -\alpha_2\frac{r_-}{\ell}
 +(\alpha_0-\alpha_2)\sqrt{-b}\right].
 \label{eq:entropy-final}
\end{equation}
Its $\alpha_0$ part, $\frac{k}{2\pi}\alpha_0\rho_a\omega^a$, comes from the exotic Lorentz Chern-Simons term, which is Lorentz invariant only up to an exact form; it is the corresponding term of the Chern-Simons surface charge~\cite{Banados1995GlobalCharges}. For the entropy of Chern-Simons terms of this kind, see also Ref.~\cite{Tachikawa2007}.

In the regular geometric representative, the complete $\alpha_2$ block coincides with the Lorentz charge of the mixed coupling:
\begin{equation}
 S_{\alpha_2}
 =-2\pi\beta\oint_{\mathcal B}\widetilde n_aA^a
 =-2\pi k\alpha_2\left(\frac{r_-}{\ell}+\sqrt{-b}\right),
 \qquad \beta=\frac{k\alpha_2}{2\pi}.
 \label{eq:adsl-wz-smooth}
\end{equation}

In the regular representative the entire $\alpha_2$ block is thus
captured by the Lorentz-only compensator acting on the mixed coupling.
The $\alpha_2$ block is unchanged when the connection and the lift are
transformed together by gauge transformations that are independent of
the angle on the evaluation circle, which include those of
Appendix~\ref{app:regular-equivalence}. The isolated $A\wedge R$ term is not
invariant under the full AdS-Lorentz group, being completed by the
remaining $\alpha_2$ terms of Eq.~\eqref{eq:adsl-full-action}. However,
for a Lorentz-valued lift these terms contribute no charge on
$\mathcal B$, so in any representative regular at $\mathcal B$ with such
a lift the whole $\alpha_2$ block is carried by the mixed coupling. The
BMS representative is not of this type: its fields cannot be extended
smoothly to $\mathcal B$ (Eq.~\eqref{eq:bms-bifurcation-obstruction}),
its thermal lift has $P$ and $Z$ components, and the Lorentz part of the
isolated mixed term on the stationary circle $\mathcal C_H$ gives
$\pi k\alpha_2(y^2-q^2)/q$ instead (Appendix~\ref{app:bms-route}).
In particular, the whole
$\sqrt{-b}$ term cannot be assigned to the mixed coupling: setting
$\alpha_2=0$ leaves $2\pi k\alpha_0\sqrt{-b}$ from the exotic sector.

For $8G_N=1$, Eq.~\eqref{eq:entropy-final} becomes
\begin{equation}
 S=4\pi\left[\alpha_1r_+-\alpha_2r_-
 +(\alpha_0-\alpha_2)\ell\sqrt{-b}\right].
 \label{eq:adsl-final-entropy}
\end{equation}

On the same $s_0=+1$ branch, Eq.~\eqref{eq:adsl-final-entropy} agrees with Eq.~(3.13) of Ref.~\cite{AvilesHidalgoValdivia2023}.\footnote{Substituting $\mathcal L_0^\pm=(\ell M\mp J)/2$ and $\mathcal L_0=-\ell b/2$, with $M=x^2+y^2$ and $J=2\ell xy$ as defined in Ref.~\cite{AvilesHidalgoValdivia2023}, into their Eq.~(3.13) reproduces Eq.~\eqref{eq:adsl-final-entropy}. The independent-sector contribution in their final entropy expression carries an additional factor of two that does not follow from their Eq.~(3.13). This factor is erroneous, and without it their final result coincides with ours.} Our construction realizes this charge explicitly using fields smooth at $\mathcal B$. For $\alpha_0<\alpha_2$, see the discussion of the $s_0$ branch in Appendix~\ref{app:s0-branch}.

\section{Discussion and Outlook}

The mixed gauge-curvature coupling contributes the horizon boost charge
$S_{\rm WZ}=-2\pi\beta\oint_{\mathcal B}\widetilde n_aA^a$ whenever the
fields and the selected geometric lift are regular at the bifurcation
circle. Its linear dependence on the projected connection follows
directly from the momentum conjugate to Lorentz curvature. In the
Maxwell case, $D_{\mathcal B}\widetilde n=0$ makes this circulation
invariant under single-valued $Z$ shifts and gives it the reduced
Abelian holonomy interpretation described above (Appendix~\ref{app:abelian-connection}).

For the AdS-Lorentz black hole, an independent-sector gauge transformation changes torsion while preserving the BTZ metric and asymptotic data, yielding a smooth bifurcation-surface representative in which the full entropy is the geometric boost charge with normalization of the independent-sector lift and the branch $s_0$ as inputs. The Euler-current comparison remains a matching of local curvature responses, since the absence of a smooth stationary unit timelike vector at the bifurcation circle obstructs its interpretation as an equilibrium entropy.

An alternative interpretation could be developed using the conical method~\cite{BTZ1994,CarlipTeitelboim1995,FursaevSolodukhin1995}.
An infinitesimal Euclidean angular deficit introduces curvature
supported on the bifurcation circle, so a term linear in curvature
can generate a defect action proportional to the same projected
connection integral. Establishing this correspondence for the full
AdS-Lorentz theory requires a specified continuation of the regular
fields, boundary terms and thermal lift. It would connect the boost
response to the curvature-bound charge familiar in quantum Hall
systems without identifying the two physical charges.

The holonomy interpretation discussed here should be distinguished from the gravitational phase operator introduced by Anandan, which describes the transport of a quantum probe through the coframe and Lorentz
connection~\cite{Anandan1996}. The Wald entropy related to the relativistic Wen-Zee term instead measures
the horizon boost charge induced by the mixed coupling and involves
the projected connection of the additional Maxwell sector.
Both constructions involve spacetime gauge geometry, but identifying
the entropy contribution with a probe's gravitational interference
phase would require specifying its coupling to the extended gauge
connection and the corresponding charge representation.

A further direction is provided by enlarging the underlying gauge algebra. Three-dimensional conformal gravity admits a Chern-Simons formulation based on the conformal group $SO(3,2)$ \cite{HorneWitten1989}. It would be interesting to determine whether Maxwell-like extensions of the conformal algebra admit invariant tensors generating analogous mixed gauge-curvature couplings and, if so, whether their horizon charges possess a corresponding holonomy interpretation. Higher-spin extensions offer a related setting in which the additional gauge sectors carry genuinely higher-rank charges \cite{Ammon2013,Kraus2013}. These generalizations would test how broadly the relation between mixed gauge-curvature couplings, horizon surface charges, and global gauge data extends within three-dimensional Chern-Simons gravity.

\acknowledgments
We thank Kartik Prabhu and Bindusar Sahoo for insightful discussions. We thank L.~Avil\'es for correspondence confirming the typographical error in Ref.~\cite{AvilesHidalgoValdivia2023}.
SH and AB were supported by ANRF Prime
Minister Early Career Grant(PM-ECRG):\\ ANRF/ECRG/2024/002083/PMS.
GP acknowledges support from national funds by FCT - Fundação para a Ciência e Tecnologia, I.P. in the framework of the project UID/04564/2025, with DOI identifier 10.54499/UID/04564/2025.

\paragraph{Use of AI-assisted tools.}
During the preparation of this work, the authors used ChatGPT
(OpenAI) and Claude (Anthropic) to assist with drafting
and revising parts of the manuscript, literature searches, algebraic
and conceptual consistency checks, and LaTeX preparation.
The authors reviewed and revised the AI-assisted material and take
full responsibility for the scientific arguments, calculations,
interpretations, references and final content of the manuscript.

\appendix

\section{Conventions and an algebraic identity}
\label{app:conventions}

Frame indices are $a,b,c=0,1,2$, spacetime indices $\mu,\nu,\lambda$,
and the coordinates $(t,r,\phi)$.  In the first-order formulation the
dreibein $e^a$ and the dualized spin connection
\begin{equation}
 \omega^a=\frac12\epsilon^{abc}\omega_{bc}
\end{equation}
are independent one-forms, and the Lorentz-covariant derivative of a
Lorentz-vector-valued form is
\begin{equation}
 D\Phi^a=\mathrm{d}\Phi^a+\epsilon^a{}_{bc}\,\omega^b\wedge\Phi^c .
\end{equation}
Throughout we abbreviate
\begin{equation}
 (X\times Y)^a\equiv\epsilon^a{}_{bc}X^b\wedge Y^c,
 \qquad
 X\cdot Y\equiv\eta_{ab}X^aY^b,
 \qquad
 \|X\|\equiv\sqrt{X\cdot X},
 \label{eq:cross-product}
\end{equation}
the wedge being understood when one entry is a zero-form.

The following elementary fact about
$\mathfrak{sl}(2,\mathbb R)\cong\mathbb R^{1,2}$ is used repeatedly.
For nonzero Lie-algebra elements (zero-forms) $X$,
\begin{equation}
 X\times Y=0
 \quad\Longleftrightarrow\quad
 Y\parallel X,
 \qquad\text{i.e.}\quad
 \ker(\operatorname{ad}_X)=\mathbb RX .
 \label{eq:centraliser}
\end{equation}
One direction is immediate from antisymmetry.  For the converse, note
that $\epsilon^a{}_{bc}X^bY^c$ is the dual of $X\wedge Y$ and that
$\eta_{ab}$ is invertible, so the two vanish together; $X\wedge Y=0$
then forces $X$ and $Y$ to be linearly dependent. 
Two equivalent tangent-space bases are used:
\begin{equation}
 \text{(i)}\quad \eta_{ab}=\operatorname{diag}(-1,1,1),
 \qquad
 \text{(ii)}\quad \eta_{01}=\eta_{10}=\eta_{22}=1,
 \label{eq:etaconventions-app}
\end{equation}
all other components in (ii) vanishing, so that
$V_aU^a=V^0U^1+V^1U^0+V^2U^2$.  The second is the null basis of
Ref.~\cite{Aviles2026}; general expressions are independent of the
choice, but explicit evaluations use the null basis to permit direct
comparison with that reference.  The Levi-Civita tensor is normalized
by $\epsilon_{012}=+1$, whence $\epsilon^{012}=-1$ (since
$\det\eta=-1$), in either basis.  The spacetime Levi-Civita tensor is fixed consistently with the chosen orientation and metric; an alternating coordinate symbol such as $[012]=+1$ is a density and must be distinguished from a tensor with indices raised by the Lorentzian metric. The Euler-current response contains two Levi-Civita tensors and is independent of their common orientation reversal.

The two bases in Eq.~\eqref{eq:etaconventions} are related by
$V^0=(\tilde e^0+\tilde e^1)/\sqrt2$, $V^1=(-\tilde e^0+\tilde e^1)/\sqrt2$,
$V^2=\tilde e^2$, with $\tilde e^a$ the orthonormal triad; this
transformation has unit determinant, so $\epsilon_{abc}$ is common to both
bases. We note that, interchanging the two null legs leaves $\eta_{01}=\eta_{10}=\eta_{22}=1$ and $g_{\phi\phi}=2e^0{}_\phi e^1{}_\phi+(e^2{}_\phi)^2$ unchanged, but reverses the orientation and hence the sign of $\epsilon_{abc}$.

The angular direction is represented differently in the two frames. In a diagonal frame adapted to a circular horizon, it may be carried by \(e^2\), whereas the particular Maxwell horizon frame of Eq.~\eqref{eq:horizondata} has \(e^2{}_\phi=0\), with the angular direction instead represented in the \((e^0,e^1)\) components. This is a property of the chosen frame, not of the null basis itself. In the regular AdS–Lorentz frame, by contrast, \(e^2{}_\phi=r_+\). Accordingly, intermediate frame components of the binormal depend on the choice of basis, while the geometric contractions constructed from them do not.

\section{Metric derivation of the Euler-current entropy}
\label{app:euler-current-derivation}

The Lagrangian component contributing to the horizon entropy is
\begin{equation}
    \mathcal L_R = \frac{\kappa \sigma}{16\pi} \, A_\mu \, \varepsilon^{\mu\nu\rho} \varepsilon^{\alpha\beta\gamma} u_\alpha R_{\nu\rho\beta\gamma}.
    \label{eq:euler-lagrangian-app}
\end{equation}
Applying the Wald formula for a stationary circular horizon $\mathcal{B}$,
\begin{equation}
    S = -2\pi \oint_{\mathcal{B}} \frac{\partial\mathcal L}{\partial R_{abcd}} \, \varepsilon^\perp_{ab}\varepsilon^\perp_{cd} \, \sqrt{h} \; \mathrm{d}\phi,
    \qquad
    \sqrt h\big|_{\mathcal{B}} = \sqrt{g_{\phi\phi}}\big|_{\mathcal{B}}
    \label{eq:euler-wald-metric}
\end{equation}
where the normal binormal is normalized as $\varepsilon^\perp_{ab}\varepsilon^{\perp ab} = -2$.

Contracting the variation with the binormals yields
\begin{equation}
    \frac{\partial\mathcal L}{\partial R_{abcd}} \varepsilon^\perp_{ab}\varepsilon^\perp_{cd} = \frac{\kappa\sigma}{16\pi} \, A_\mu \Big(\varepsilon^{\mu\nu\rho}\varepsilon^\perp_{\nu\rho}\Big) \Big(\varepsilon^{\alpha\beta\gamma} u_\alpha \varepsilon^\perp_{\beta\gamma}\Big).
    \label{eq:euler-derivative-contracted}
\end{equation}
In an adapted chart $(t, r, \phi)$ where $(t,r)$ span the normal plane, the binormal components are $\varepsilon^\perp_{tr} = -\varepsilon^\perp_{rt} = s_\perp \sqrt{-g}/\sqrt{g_{\phi\phi}}$ with orientation choice $s_\perp = \pm 1$. Letting $o = \pm 1$ denote the Levi-Civita density orientation, the two contractions reduce to
\begin{equation}
    \varepsilon^{\mu\nu\rho}\varepsilon^\perp_{\nu\rho} = \frac{2\,o\,s_\perp\,\delta^\mu_\phi}{\sqrt{g_{\phi\phi}}},
    \qquad
    \varepsilon^{\alpha\beta\gamma}u_\alpha \varepsilon^\perp_{\beta\gamma} = \frac{2\,o\,s_\perp\,u_\phi}{\sqrt{g_{\phi\phi}}}.
\end{equation}
Substituting these back into Eq.~\eqref{eq:euler-derivative-contracted}, the orientation factors square to unity ($o^2 = s_\perp^2 = 1$), leaving
\begin{equation}
    \frac{\partial\mathcal L}{\partial R_{abcd}} \varepsilon^\perp_{ab}\varepsilon^\perp_{cd} = \frac{\kappa\sigma \, A_\phi u_\phi}{4\pi \, g_{\phi\phi}}.
\end{equation}
Evaluating the surface integral \eqref{eq:euler-wald-metric} with $\sqrt{h} = \sqrt{g_{\phi\phi}}$ gives
\begin{equation}
    \Delta S_{\rm WZ} = -\frac{\kappa\sigma}{2} \oint_{\mathcal{B}} \frac{A_\phi u_\phi}{\sqrt{g_{\phi\phi}}} \, \mathrm{d}\phi 
\end{equation}
Expressing this in tangent form, with $\hat t^\mu=\delta^\mu_\phi/\sqrt{g_{\phi\phi}}$ and
$\mathrm{d} s=\sqrt{g_{\phi\phi}}\,\mathrm{d}\phi$, we get
\begin{equation}
 \frac{A_\phi u_\phi}{\sqrt{g_{\phi\phi}}}\,\mathrm{d}\phi
 =\frac{A_\phi u_\phi}{g_{\phi\phi}}\,\mathrm{d} s
 =\big(A\cdot\hat t\big)\big(u\cdot\hat t\big)\,\mathrm{d} s,
\end{equation}
which is the formal curvature response of Eq.~\eqref{eq:eulerentropyinvariantform}, subject to the timelike-vector obstruction discussed in the main text.

\section{Derivation of the Wen-Zee Noether charge}
\label{app:noether}

This appendix derives the general horizon charge of
\begin{equation}
 L_{\rm WZ}=\beta A^a\wedge R_a
\end{equation}
in first-order variables.  It applies when the Lorentz-only compensator
is consistent, in particular on the torsion-free Maxwell backgrounds of
the main calculation. In the regular AdS-Lorentz representative this potential reproduces the complete $\alpha_2$ contribution, Eq.~\eqref{eq:adsl-wz-smooth}. Its interpretation as a separate mixed-term contribution is restricted to that geometric representative.

\subsection{Variation and pre-symplectic potential}

Since $\delta R^a=D\delta\omega^a$, the variation is
\begin{equation}
 \delta L_{\rm WZ}
 =\beta\,\delta A^a\wedge R_a
 +\beta A^a\wedge D\delta\omega_a.
\end{equation}
For a Lorentz-vector-valued one-form $A^a$, covariant integration by
parts gives
\begin{equation}
 A^a\wedge D\delta\omega_a
 =DA^a\wedge\delta\omega_a
 -d\left(A^a\wedge\delta\omega_a\right).
\end{equation}
Therefore
\begin{equation}
 \delta L_{\rm WZ}
 =\beta R_a\wedge\delta A^a
 +\beta\delta\omega_a\wedge DA^a
 +d\Theta(\delta),
 \qquad
 \Theta(\delta)=-\beta A^a\wedge\delta\omega_a.
 \label{eq:appC:theta}
\end{equation}
The corresponding Euler-Lagrange forms contributed by this term are
$E_{A,a}=\beta R_a$ and $E_\omega{}^a=\beta DA^a$.

\subsection{Local-Lorentz charge}

Under an infinitesimal Lorentz rotation with parameter $\rho^a$,
\begin{equation}
 \delta_\rho\omega^a=D\rho^a,
 \qquad
 \delta_\rho A^a=\epsilon^a{}_{bc}A^b\rho^c.
\end{equation}
Because $A^a\wedge R_a$ is exactly Lorentz invariant, its Noether
current contains no additional boundary term:
\begin{equation}
 J_\rho=\Theta(\delta_\rho)
 =-\beta A^a\wedge D\rho_a.
\end{equation}
Using
\begin{equation}
 A^a\wedge D\rho_a
 =\rho_aDA^a-d(\rho^aA_a),
\end{equation}
we obtain
\begin{equation}
 J_\rho=d\left(\beta\rho^aA_a\right)-\rho_aE_\omega{}^a,
 \qquad
 Q_\rho=\beta\rho^aA_a.
 \label{eq:appC:Qrho}
\end{equation}

\subsection{The combined Lorentz-diffeomorphism charge}

For a vector field $\xi$, define
$\lambda^a\equiv\iota_\xi\omega^a$.  Cartan's identity gives
\begin{equation}
 \mathcal{L}_\xi\omega^a=\iota_\xi R^a+D\lambda^a.
\end{equation}
Substitution into Eq.~\eqref{eq:appC:theta} yields the
pure-diffeomorphism charge
\begin{equation}
 Q^{\rm diff}_\xi=\beta(\iota_\xi\omega^a)A_a.
\end{equation}
For a smooth connection it vanishes on the bifurcation surface because $\xi|_\mathcal B=0$ and is
not separately Lorentz covariant.  The relevant symmetry is the
combined Lorentz-diffeomorphism as shown in \cite{JacobsonMohd},
\begin{equation}
 \widehat\delta=\mathcal{L}_\xi+\delta_\rho,
 \qquad
 Q_{(\xi,\rho)}
 =\beta\left(\iota_\xi\omega^a+\rho^a\right)A_a.
 \label{eq:appC:combined-app}
\end{equation}
On $\mathcal B$ only the Lorentz contribution survives.

\subsection{Fixing the compensating Lorentz parameter}
\label{sec:fixinglorentzcompensatormaxwell}

To evaluate this charge at the bifurcation surface
$\mathcal B$, we require the geometric properties of the
horizon.  In $2+1$ dimensions, the bifurcation surface is a
one-dimensional closed curve.  The standard Iyer-Wald relation fixes
the derivative of the Killing field as
\begin{equation}
 \nabla^a\xi^b\big|_{\mathcal B}=\kappa_{\rm sg} n^{ab},
 \qquad n_{ab}n^{ab}=-2,
\end{equation}
where $n_{ab}$ is the horizon binormal.  Taking the dual gives
\begin{equation}
 \widetilde n^a=\frac12\epsilon^{abc}n_{bc}.
\end{equation}

Because $\widetilde n^a$ is orthogonal to the two-dimensional normal
plane of $\mathcal B$, it must point along the one-dimensional curve
itself.  It is therefore the unit tangent to the bifurcation circle,
opposite to $+\partial_\phi$ for the angular orientation chosen in the main text, with
\begin{equation}
 \widetilde n^a=\frac{-\,e^a{}_\phi}{\sqrt{g_{\phi\phi}}},
 \qquad
 \eta_{ab}\widetilde n^a\widetilde n^b=1 .
\end{equation}
The pullback to $\mathcal B$ has only an angular component. A displacement along the curve is
$\mathrm{d} x^\mu=\hat t^\mu\mathrm{d} s$ with
$\hat t^\mu=\delta^\mu{}_\phi/\sqrt{g_{\phi\phi}}$.  Evaluating $e^a$ on that
displacement gives $e^a{}_\mu\hat t^\mu=-\widetilde n^a$, hence
\begin{equation}
 e^a\big|_{\mathcal B}=-\,\widetilde n^a\,\mathrm{d} s .
 \label{eq:appC:pullback}
\end{equation}

The compensating rotation is fixed by demanding frame invariance,
$\widehat\delta e^a=0$.  Introduce
$\xi^a\equiv\iota_\xi e^a$.  The Lie derivative of the frame can be
written covariantly as
\begin{equation}
 \mathcal{L}_\xi e^a
 =\iota_\xi T^a+D\xi^a
 -\epsilon^a{}_{bc}\lambda^be^c.
\end{equation}
Since
$\delta_\rho e^a=-\epsilon^a{}_{bc}\rho^be^c$, the condition
$\widehat\delta e^a=0$ becomes
\begin{equation}
 \iota_\xi T^a+D\xi^a
 -\epsilon^a{}_{bc}(\lambda^b+\rho^b)e^c=0.
 \label{eq:appC:framecondition}
\end{equation}
At the bifurcation surface, $\xi^\mu=0$, and hence
$\xi^a=\lambda^a=\iota_\xi T^a=0$.  Equation
\eqref{eq:appC:framecondition} therefore reduces to
\begin{equation}
 \nabla_c\xi^a=\epsilon^a{}_{bc}\rho^b.
 \label{eq:appC:rhocond}
\end{equation}
The Iyer-Wald relation is $\nabla^a\xi^b|_{\mathcal B}=\kappa_{\rm sg}n^{ab}$, whereas the frame-invariance condition involves $\nabla_c\xi^a$. Using the Killing identity, $\nabla_c\xi^a=-\nabla^a\xi_c=-\kappa_{\rm sg}n^a{}_c$, Eq.~\eqref{eq:appC:rhocond} becomes $-\kappa_{\rm sg}n^a{}_c=\epsilon^a{}_{bc}\rho^b.$ Contracting with $\epsilon_a{}^{dc}$ and using $\epsilon_a{}^{dc}\epsilon^a{}_{bc}=-2\delta^d_b$ and $\epsilon_a{}^{dc}n^a{}_c=-2\widetilde n^d$ gives $2\kappa_{\rm sg}\widetilde n^d=-2\rho^d$, and hence frame invariance fixes the internal rotation parameter uniquely:
\begin{equation}
    \label{eq:appC:rho}
    \rho^a=-\kappa_{\rm sg}\widetilde n^a.
\end{equation}
This is consistent with the Lorentz-diffeomorphism Noether-charge convention of Ref.~\cite{JacobsonMohd}, which reproduces the standard Iyer-Wald entropy expression.

\subsection{Entropy}

Wald's relation on the bifurcation surface is
\begin{equation}
 S=\frac{2\pi}{\kappa_{\rm sg}}\oint_\mathcal B Q_{(\xi,\rho)}.
\end{equation}
Using Eqs.~\eqref{eq:appC:combined-app} and \eqref{eq:appC:rho}, the surface
gravity cancels and one obtains
\begin{equation}
 S_{\rm WZ}=-2\pi\beta\oint_\mathcal B\widetilde n^aA_a.
 \label{eq:appC:entropy}
\end{equation}
Using Eq.~\eqref{eq:appC:pullback} gives the projected line integral in the main text.

\section{Abelian transformation of the projected field under \texorpdfstring{$Z$}{Z} shifts}
\label{app:abelian-connection}

In this appendix we show that, in the Maxwell sector, the projected field
$a_{\mathcal B}=\widetilde n_aA^a|_{\mathcal B}$ transforms as an Abelian
connection under $Z$ shifts. This underlies the holonomy interpretation of
Sec.~\ref{sec:holonomy}. All fields and covariant derivatives below are
pulled back to $\mathcal B$.

For the Maxwell algebra $[Z_a,P_b]=[Z_a,Z_b]=0$, so a bare $Z$-shift
with parameter $\zeta^a$ leaves the vielbein and spin connection invariant,
\begin{equation}
 \delta_\zeta e^a=0,
 \qquad
 \delta_\zeta\omega^a=0,
 \label{eq:appD:Zshift-frame-invariance}
\end{equation}
independently of any background or equation of motion.  Thus a Maxwell $Z$ shift leaves both the metric and the projection direction unchanged.

The $Z$-sector transformation acts on the gauge field as
\begin{equation}
 \delta_\zeta A^a=D\zeta^a.
\end{equation}
On the torsion-free Maxwell background $D=\mathring D$. Since
$\mathcal B$ is a closed geodesic with unit tangent $\widetilde n$,
$\mathring D\widetilde n^a|_{\mathcal B}=0$, which gives
\begin{equation}
 D\widetilde n^a\big|_{\mathcal B}=0.
 \label{eq:appD:Dn}
\end{equation}

The projected field therefore transforms as
\begin{equation}
 \delta_\zeta a_{\mathcal B}
 =\widetilde n_aD\zeta^a
 =d(\widetilde n_a\zeta^a)-\zeta^aD\widetilde n_a
 =d_{\mathcal B}(\widetilde n\cdot\zeta),
 \label{eq:appD:abelian-law}
\end{equation}
where the last step uses Eq.~\eqref{eq:appD:Dn}. This is the
transformation law of an Abelian connection on $\mathcal B$, with gauge
parameter $\widetilde n\cdot\zeta$. Consequently $\oint_{\mathcal B}a_{\mathcal B}$,
and with it the entropy contribution
$S_{\rm WZ}=-2\pi\beta\oint_{\mathcal B}a_{\mathcal B}$, is invariant under
single-valued $Z$ shifts. Transformations with nontrivial winding require a specified global gauge group and charge representation; no such large-gauge identification is assumed here. If $A^a$ is considered on a background with torsion, $D\widetilde n^a|_{\mathcal B}=(K\times\widetilde n)^a|_{\mathcal B}$, and the same transformation law holds whenever the pulled-back contorsion is aligned with $\widetilde n$.

\section{Evaluation on the Cosmological Solution}
\label{app:fsc-evaluation}

This appendix exhibits a representative of the flat cosmological solution that is smooth at the bifurcation circle, on which the Lorentz compensator is an exact symmetry of $e$, $\omega$ and $A$. Its angular data at $\mathcal B$ are a constant Lorentz rotation of Eq.~\eqref{eq:horizondata}, which justifies evaluating the
Wald charge of Sec.~\ref{sec:entropy:maxwell-comparison} with those data.
Let $r_0=r_c$, $v_0=\operatorname{sgn}(\mathcal N)\sqrt{\mathcal M}$,
$c=\Xi/(2v_0r_0)$ and $z=(0,0,1)$. On Minkowski space
$\mathrm{d} s^2=\mathrm{d} U\mathrm{d} V+\mathrm{d} Y^2$, impose
$(U,V,Y)\sim(e^{-2\pi v_0}U,e^{2\pi v_0}V,Y+2\pi r_0)$.
With $X_\perp=(V,U,0)/\sqrt2$ and $L=\exp[-(v_0Y/r_0)z\times]$, a smooth quotient-invariant solution is
\begin{equation}
 \begin{split}
 e&=L\left(\frac{\mathrm{d} V}{\sqrt2},\frac{\mathrm{d} U}{\sqrt2},\mathrm{d} Y\right),
 \qquad \omega=\frac{v_0}{r_0}z\,\mathrm{d} Y,\\
 A&=L\left(-\tfrac12X_\perp\times\mathrm{d} X_\perp
              -X_\perp\times z\,\mathrm{d} Y\right)+cz\,\mathrm{d} Y.
 \end{split}
\end{equation}
Direct differentiation gives $R=T=0$ and
$D_\omega A+\tfrac12e\times e=0$. The boost
$\xi=\kappa_{\rm sg}(V\partial_V-U\partial_U)$ fixes
$\mathcal B=\{U=V=0\}$ and has the compensator
$\rho=\kappa_{\rm sg}z=-\kappa_{\rm sg}\widetilde n$.
Writing $\phi=Y/r_0$, the horizon fields are
$e_\phi=r_0z$, $\omega_\phi=v_0z$, $A_\phi=cr_0z$.
Their invariants are
\begin{equation}
 \mathcal M=v_0^2,\quad \mathcal N=2v_0r_0,\quad
 \mathcal F=r_0^2+2v_0cr_0,\quad \Xi=2v_0cr_0.
\end{equation}
They are related by a constant Lorentz rotation to the collinear data
in Eq.~\eqref{eq:horizondata}. This proves the existence of smooth
fields realizing those pullbacks and fixes the orientation used below.
For the same geometric lift, the full CS pairing is
$k\oint z_a(\alpha_0\omega^a+\alpha_1e^a+\alpha_2A^a)
=2\pi k(\alpha_0v_0+\alpha_1r_0+\alpha_2cr_0)$,
which gives the full entropy quoted in the main text.
Since $z^aZ_a$ is covariantly constant and, by $[P_a,Z_b]=0$, acts only on $A$,
adding it to the lift generates an internal symmetry and adds the corresponding
$Z$ charge; it does not change the boost charge, and the geometric lift has no
$Z$ part at $\mathcal B$. The AdS-Lorentz case differs here. Its free direction
$\widetilde n^a(J_a-Z_a)$ also leaves $e$ unchanged, since $[J_a-Z_a,P_b]=0$, but
it rotates the independent connection $\omega$, whose rate of rotation under the
horizon boost the geometry does not fix; this is why the normalization in
Sec.~\ref{sec:adsl-lift} is an input there, whereas here $\omega$ is the
Levi-Civita connection of $e$.

After identifying the gauge field with the on-shell gravitational Maxwell field, the horizon data in
Eq.~\eqref{eq:horizondata} are
\begin{equation}
 e^a{}_{\phi}\big|_{r_c}
 =\left(\frac{\mathcal N}{4},
 \frac{\mathcal N}{2\mathcal M},0\right),
 \qquad
 A^a{}_{\phi}\big|_{r_c}
 =\left(\frac{\Xi}{4},
 \frac{\Xi}{2\mathcal M},0\right),
 \qquad
 \Xi=\mathcal F-\frac{\mathcal N^2}{4\mathcal M}.
 \label{eq:appFSC:horizondata}
\end{equation}

In the null basis,
$\eta_{01}=\eta_{10}=\eta_{22}=1$, so the induced metric on the
horizon circle is
\begin{align}
 g_{\phi\phi}
 &=\eta_{ab}e^a{}_{\phi}e^b{}_{\phi}
 =2e^0{}_{\phi}e^1{}_{\phi}
 +(e^2{}_{\phi})^2 \notag\\
 &=2\left(\frac{\mathcal N}{4}\right)
 \left(\frac{\mathcal N}{2\mathcal M}\right)
 =\frac{\mathcal N^2}{4\mathcal M},
 \qquad
 \sqrt{g_{\phi\phi}}
 =\frac{|\mathcal N|}{2\sqrt{\mathcal M}}.
 \label{eq:appFSC:metric}
\end{align}
The dualized horizon binormal, opposite to the unit tangent for the
orientation used in the main text, is therefore
\begin{align}
 \widetilde n^a
 &=-\frac{e^a{}_{\phi}}{\sqrt{g_{\phi\phi}}} \notag\\
 &=-\operatorname{sgn}(\mathcal N)
 \left(\frac{\sqrt{\mathcal M}}{2},
 \frac{1}{\sqrt{\mathcal M}},0\right).
 \label{eq:appFSC:tangent}
\end{align}

Lowering the index on the Maxwell field interchanges its first two
components because of the null basis,
\begin{equation}
 A_{0\phi}=A^1{}_{\phi}
 =\frac{\Xi}{2\mathcal M},
 \qquad
 A_{1\phi}=A^0{}_{\phi}
 =\frac{\Xi}{4},
 \qquad
 A_{2\phi}=0.
 \label{eq:appFSC:lowered}
\end{equation}
The projected contraction is consequently
\begin{align}
 \widetilde n^aA_{a\phi}
 &=-\operatorname{sgn}(\mathcal N)
 \left[
 \frac{\sqrt{\mathcal M}}{2}
 \frac{\Xi}{2\mathcal M}
 +\frac{1}{\sqrt{\mathcal M}}\frac{\Xi}{4}
 \right] \notag\\
 &=-\operatorname{sgn}(\mathcal N)
 \frac{\Xi}{2\sqrt{\mathcal M}} \notag\\
 &=-\operatorname{sgn}(\mathcal N)\,
 \frac12\left(
 \frac{\mathcal F}{\sqrt{\mathcal M}}
 -\frac{\mathcal N^2}{4\mathcal M^{3/2}}
 \right),
 \label{eq:appFSC:projection}
\end{align}
which is the projected holonomy used in the torsion-free Maxwell limit.

Finally, substituting Eq.~\eqref{eq:appFSC:projection} into the 
projected-holonomy formula Eq.~\eqref{eq:WZ-projected} with $\beta=k\alpha_2/2\pi$ gives
\begin{equation}
 S
 =\pi k \alpha_2
 \operatorname{sgn}(\mathcal N)
 \left(
 \frac{\mathcal F}{\sqrt{\mathcal M}}
 -\frac{\mathcal N^2}{4\mathcal M^{3/2}}
 \right).
 \label{eq:FSCresultappendix}
\end{equation}
This reproduces \eqref{eq:FSCresult} and hence matches
the entropy of Ref.~\cite{Aviles2026}.

\section{Regular AdS-Lorentz fields near the bifurcation circle}
\label{app:kruskal}

For the metric~\eqref{eq:adsl-btz-metric}, define
\begin{equation}
 G(r)=\frac{r-r_+}{r+r_+}
 \left(\frac{r-r_-}{r+r_-}\right)^{-r_-/r_+},\qquad
 r_* =\frac{\log G}{2\kappa_{\rm sg}},\qquad UV=-G(r)
 \label{eq:kruskal-G}
\end{equation}
in the exterior. Differentiation gives $r_*'=N^{-2}$.
Since $G(r_+)=0$ and
\begin{equation}
 G'(r_+)=\frac1{2r_+}
 \left(\frac{r_+-r_-}{r_++r_-}\right)^{-r_-/r_+}\ne0,
\end{equation}
$r$ is analytic in $UV$ near the outer horizon. Set
\begin{equation}
 \mathsf A=\frac{N^2}{\kappa_{\rm sg}^2G},\qquad
 \mathsf H=-\frac{\Omega_H+N^\varphi}{2\kappa_{\rm sg}G}.
 \label{eq:kruskal-AH}
\end{equation}
The metric becomes
\begin{equation}
 \mathrm{d} s^2=-\mathsf A\mathrm{d} U\mathrm{d} V+
 r^2\big[\mathrm{d}\widetilde\phi+\mathsf H(U\mathrm{d} V-V\mathrm{d} U)\big]^2.
\end{equation}
The apparent singularities in~\eqref{eq:kruskal-AH} are removable.
Writing $f=((x-y)/(x+y))^{y/x}$, their limits are
\begin{equation}
 \mathsf A(0)=\frac{4\ell^2x^2}{x^2-y^2}f>0,\qquad
 \mathsf H(0)=-\frac{2y}{x^2-y^2}f.
\end{equation}
Thus~\eqref{eq:adsl-regular-frame} is a smooth, nondegenerate frame
across both branches of the outer horizon. Under the Killing flow,
\begin{equation}
 \mathcal{L}_\xi e^0=\kappa_{\rm sg}e^0,\qquad
 \mathcal{L}_\xi e^1=-\kappa_{\rm sg}e^1,\qquad
 \mathcal{L}_\xi e^2=0.
\end{equation}
This is the constant boost generated by $\rho=(0,0,\kappa_{\rm sg})$.
Direct evaluation of $n_{ab}=e_a{}^\mu e_b{}^\nu
\nabla_\mu\xi_\nu/\kappa_{\rm sg}$ at $\mathcal B$ gives
$n_{01}=1$ and $\widetilde n=(0,0,-1)$.

Solving $\mathrm{d} e+\mathring{\omega}\times e=0$ at $\mathcal B$ yields
\begin{equation}
 \begin{split}
 \mathring{\omega}_U&=\left(0,-\frac{\sqrt2\mathsf H(0)r_+}{\sqrt{\mathsf A(0)}},0\right),\\
 \mathring{\omega}_V&=\left(\frac{\sqrt2\mathsf H(0)r_+}{\sqrt{\mathsf A(0)}},0,0\right),\qquad
 \mathring{\omega}_\phi=\left(0,0,\frac{2\mathsf H(0)r_+^2}{\mathsf A(0)}\right)
 =(0,0,-y).
 \end{split}
\end{equation}
Together with $e_\phi=(0,0,r_+)$ and
$\omega=-s_0q\widetilde n\mathrm{d}\widetilde\phi$, these give
Eq.~\eqref{eq:adsl-smooth-pullbacks}; the pullbacks are assembled in
Appendix~\ref{app:regular-pullbacks}. In particular
$K_\phi=-(y+s_0q)\widetilde n$ is regular and aligned with the boost axis.

\subsection{Exterior equivalence and asymptotic conditions}
\label{app:regular-equivalence}
The exterior solution of Ref.~\cite{AvilesHidalgoValdivia2023} contains an independent flat Lorentz connection. In the BMS representative used there, it reads
\begin{equation}
 \omega_{\rm ext}=X\,\mathrm{d}\phi_{\rm BMS},
 \qquad
 X^a=\left(\frac{q^2}{2},1,0\right),
 \qquad
 X^2=q^2,
\end{equation}
where we have used \(q^2=-b\) in our conventions.

Thus \(q^2=X^2\) is the invariant magnitude of the independent flat-connection sector. We choose the regular representative such that the corresponding connection vector is aligned with the horizon dual-binormal direction \(\widetilde n\).

Since $\widetilde n^2=1$ and Lorentz transformations preserve the
invariant norm, there exists a constant $R\in SO_0(2,1)$ such that
\begin{equation}
 R^{-1}XR=-s_0q\,\widetilde n,
 \qquad s_0=\pm1.
\end{equation}
The two signs correspond to the two possible orientations of the
connection vector relative to the chosen horizon boost direction. Thus
the magnitude in~\eqref{eq:adsl-regular-independent} is not an
additional assumption: it follows from the invariant $X^2=q^2$, while
the sign remains discrete data.

Distinguish the BMS coordinates $(u,r,\phi_{\rm BMS})$ from the BTZ coordinates $(t,r,\varphi)$ used in Eq.~\eqref{eq:adsl-btz-metric}. On the overlap,
\begin{equation}
 u=t-r_*,\qquad \phi_{\rm BMS}=\varphi+g(r),\qquad
 \widetilde\phi=\varphi-\Omega_Ht,\qquad
 g'(r)=\frac{N^\varphi}{N^2}.
\end{equation}
This sign of $g'$ removes the radial-angular cross term in the BTZ metric. Thus $\phi_{\rm BMS}=\widetilde\phi+F(t,r)$, with $F=\Omega_Ht+g(r)$. The transformation
\begin{equation}
 h=e^{-FX}R,\qquad
 h^{-1}(\mathrm{d}+\omega_{\rm ext})h
 =-s_0q\widetilde n\,\mathrm{d}\widetilde\phi
\end{equation}
relates the exterior independent connection to Eq.~\eqref{eq:adsl-regular-independent}.

The exterior coframe and~\eqref{eq:adsl-regular-frame} give the same
metric, orientation and time orientation and are related by a
$\phi$-independent Lorentz rotation $g_L(t,r)$ on their overlap.
Applying $(h,g_L,g_L)$ in the three factors gives the full regular
connection. These transformations are single-valued around the angular
circle. They need not extend to $\mathcal B$ in the exterior section;
the transformed fields themselves provide the extension.

The transformation can be chosen to interpolate to the identity in an
exterior annulus, so that the asymptotic fields remain unchanged. Since
the transformation of the independent connection does not alter the
dreibein, the BTZ metric is unchanged, while the independent connection,
and hence the contorsion $K=\omega-\mathring{\omega}$, is changed.

To keep the asymptotic fields unchanged, choose $r_+<r_1<r_2$ and
interpolate the transformation to the identity in $r_1<r<r_2$.
Keep the two gravitational factors equal throughout:
$g=(h,g_L,g_L)$. Smooth interpolation exists on the exterior annulus,
since the transformations depend only on $(t,r)$ and lie in the
connected identity components; equivalently one can interpolate their
boost, rotation and exponential factors. Then
$e^g=g_L^{-1}eg_L$ remains invertible and preserves the metric, while
the transformed full connection remains flat. For $r>r_2$ the fields
are exactly the original asymptotic fields. The construction requires
only a neighbourhood of the outer bifurcation circle and the adjoining
exterior, not an extension through the inner horizon.

The interpolation can introduce time dependence into the connection. For example, replacing $e^{-FX}$ by $e^{-\lambda(r)FX}$ produces the smooth radial term $-\lambda'(r)FX\,\mathrm{d} r$ in the matching annulus. The metric remains stationary, and the transported compensator gives the combined exact symmetry $D_{\mathcal A^g}\Psi^g=0$. Since the transformation is the identity for $r>r_2$, it preserves the original asymptotic fields and boundary chemical potentials.

Transform the lift together with the connection:
\begin{equation}
 \mathcal A^g=g^{-1}\mathcal A g+g^{-1}\mathrm{d} g,\qquad
 \Lambda^g=g^{-1}\Lambda g-g^{-1}\mathcal{L}_\xi g,\qquad
 \Psi^g=g^{-1}\Psi g.
 \label{eq:thermal-lift-finite-law}
\end{equation}
The normalization~\eqref{eq:adsl-thermal-norms} is also preserved under
passage to the exterior representative, since the lifted symmetry
parameter transforms homogeneously as $\Psi_i^g=g_i^{-1}\Psi_i g_i$.
Transporting the regular boost back to the stationary exterior gives
\begin{equation}
 \Psi_0^{\rm ext}=\frac{s_0\kappa_{\rm sg}}q X,\qquad
 \Psi_\pm^{\rm ext}=\iota_\xi\mathcal A_\pm^{\rm ext},\qquad
 \Lambda_{\rm ext}=\mu_w X^a(J_a-Z_a),\qquad
 \mu_w=\frac{s_0\kappa_{\rm sg}}q-\Omega_H.
 \label{eq:exterior-lift}
\end{equation}
This is the thermal symmetry with the independent-sector potential
$\mu_w$. Its bulk lift need not be Lorentz-valued in every
representative; it is the geometric boost at the regular bifurcation circle.
The sign of $\mu_w+\Omega_H$ agrees with the invariant branch
label~\eqref{eq:adsl-branch-invariant}. Transforming the connection
and lift together preserves that label. Changing $s_0$ while imposing
the same geometric boost anew changes the ensemble choice, rather than
the evaluation of a fixed lifted symmetry.

\subsubsection{Invariant characterization of the \texorpdfstring{$s_0$}{s0} branch}
\label{app:s0-branch}
Since the entropy~\eqref{eq:entropy-lift-family} depends on $s_0$, we check that this sign is not a gauge artifact. The connections with $s_0=+1$ and $s_0=-1$ are both regular and lie in the same exterior gauge orbit, so the connection alone does not determine the sign. The two branches differ in the lifted symmetry. Imposing the geometric boost anew in each representative selects a different lift.

The sign can therefore be characterized invariantly only for the pair consisting of the connection and its lift. Let $p$ be a base point on the positively oriented angular circle, and let $H_0(p)=\mathcal P\exp(-\oint\omega)$ be the parallel-transport holonomy of $\omega$ based at $p$. We define $X_\Sigma(p)=-(2\pi)^{-1}\log H_0(p)$, using the real hyperbolic branch of the logarithm on which $X_\Sigma=\omega_\phi$ in the constant representative. Then
\begin{equation}
 s_0=\operatorname{sgn}\!\big[\Psi_0(p)\cdot X_\Sigma(p)\big],
 \qquad
 \Psi_0\cdot X_\Sigma=s_0\kappa_{\rm sg}q,
 \label{eq:adsl-branch-invariant}
\end{equation}
where the dot denotes contraction with $\eta_{ab}$, without the coupling $c_0$. Under a gauge transformation, $\Psi_0$ and $X_\Sigma$ are both conjugated at the base point. Their contraction, and hence $s_0$, is therefore unchanged when the connection and the lift are transformed together. Pairing $\Psi_0$ directly with $\omega_\phi$ would give the same sign only in a constant angular gauge.

The choice $s_0=+1$ made in the main text corresponds to the branch used in Ref.~\cite{AvilesHidalgoValdivia2023} for $\alpha_0>\alpha_2$. For $\alpha_0<\alpha_2$, that reference takes the opposite sign of the last term, which keeps the torsional contribution non-negative. In the exterior parametrization this branch choice is equivalently expressed by $\mu_w+\Omega_H>0$ in Eq.~\eqref{eq:exterior-lift}.

\subsection{Pullbacks of the regular fields on the bifurcation circle}
\label{app:regular-pullbacks}

We collect here the explicit evaluation of the regular fields on the outer bifurcation circle $\mathcal B$. This gives the horizon pullbacks used in Sec.~\ref{sec:adsl-entropy}. The regular coordinates $(U,V,\widetilde\phi)$ and the corresponding dreibein are
\begin{equation*}
U=-e^{-\kappa_{\rm sg}(t-r_*)},
\qquad
V=e^{\kappa_{\rm sg}(t+r_*)},
\qquad
\widetilde\phi=\varphi-\Omega_H t,
\end{equation*}
and
\begin{equation*}
e^0=-\sqrt{\frac{\mathsf A}{2}}\,\mathrm{d} V,\qquad
e^1=\sqrt{\frac{\mathsf A}{2}}\,\mathrm{d} U,\qquad
e^2=r\left[\mathrm{d}\widetilde\phi+\mathsf H(U\mathrm{d} V-V\mathrm{d} U)\right].
\end{equation*}
Here $\mathsf A$ and $\mathsf H$ are smooth functions of $UV$, and the bifurcation circle is given by
\begin{equation*}
\mathcal B:\qquad U=V=0,\qquad r=r_+.
\end{equation*}

Since $\mathrm{d} U$ and $\mathrm{d} V$ have no component along the tangent direction $\partial_{\widetilde\phi}$, while $U=V=0$ on $\mathcal B$, the pullback of the dreibein is simply
\begin{equation*}
e^0{}_{\widetilde\phi}\big|_{\mathcal B}=0,
\qquad
e^1{}_{\widetilde\phi}\big|_{\mathcal B}=0,
\qquad
e^2{}_{\widetilde\phi}\big|_{\mathcal B}=r_+.
\end{equation*}
Introducing the dimensionless coframe $\bar e=e/\ell$, together with
$x=\frac{r_+}{\ell},
\widetilde n=(0,0,-1)$,
we therefore obtain

\begin{equation}
\bar e^a{}_{\widetilde\phi}=-x\,\widetilde n^a .
\label{eq:F2-ebar}
\end{equation}

We next determine the angular component of the Levi-Civita connection. It is fixed by the torsion-free condition
\begin{equation*}
D_{\mathring{\omega}}\bar e
=
\mathrm{d}\bar e+\mathring{\omega}\times\bar e=0.
\end{equation*}
For the regular frame above, the angular component at the bifurcation circle is obtained by evaluating the torsion-free equations on the $(\widetilde\phi,U)$ and $(\widetilde\phi,V)$ directions. Since $\mathsf A$ and $\mathsf H$ are smooth functions of $UV$, their derivatives with respect to $U$ and $V$ are finite at $\mathcal B$, and the resulting connection is regular there. The angular component is
\begin{equation*}
\mathring{\omega}^0{}_{\widetilde\phi}\big|_{\mathcal B}=0,
\qquad
\mathring{\omega}^1{}_{\widetilde\phi}\big|_{\mathcal B}=0,
\qquad
\mathring{\omega}^2{}_{\widetilde\phi}\big|_{\mathcal B}
=-y,
\end{equation*}
where $y=\frac{r_-}{\ell}$.

Consequently, $\mathring{\omega}^a{}_{\widetilde\phi}=
y\,\widetilde n^a$.
Thus the value of $\mathring{\omega}_{\widetilde\phi}$ is fixed by the regular BTZ geometry; it is not an additional choice.

The independent AdS-Lorentz connection is different. Its regular representative, Eq.~\eqref{eq:adsl-regular-independent}, motivated in Appendix~\ref{app:regular-equivalence}, has the form
\begin{equation*}
\omega=-s_0q\,\widetilde n\,\mathrm{d}\widetilde\phi,
\qquad
q=\sqrt{-b}>0,
\qquad
s_0=\pm1.
\end{equation*}
Hence its pullback to the bifurcation circle is immediately
\begin{equation}
\omega^a{}_{\widetilde\phi}=-s_0q\,\widetilde n^a .
\label{eq:F2-omega}
\end{equation}
The parameter $q$ labels the independent flat-connection sector and is therefore not determined by the BTZ metric parameters $r_\pm$.

Finally, with the convention $A=\mathring{\omega}-\omega$, the angular component of the gauge field follows algebraically:
\begin{equation*}
A_{\widetilde\phi}
=
\mathring{\omega}_{\widetilde\phi}-\omega_{\widetilde\phi}
=
y\widetilde n+s_0q\,\widetilde n.
\end{equation*}
Thus
\begin{equation}
A^a{}_{\widetilde\phi}=(y+s_0q)\widetilde n^a .
\label{eq:F2-A}
\end{equation}

Collecting the four horizon pullbacks,
\begin{equation}
\bar e^a{}_{\widetilde\phi}=-x\widetilde n^a,
\qquad
\mathring{\omega}^a{}_{\widetilde\phi}=y\widetilde n^a,
\qquad
\omega^a{}_{\widetilde\phi}=-s_0q\widetilde n^a,
\qquad
A^a{}_{\widetilde\phi}=(y+s_0q)\widetilde n^a .
\label{eq:F2-collected}
\end{equation}
These are the quantities entering the evaluation of the Wald charge on the bifurcation circle in Sec.~\ref{sec:adsl-entropy}.

\section{The BMS-gauge route}
\label{app:bms-route}

The black hole of Ref.~\cite{AvilesHidalgoValdivia2023} is given in the
BMS gauge. This appendix records those fields, the obstruction to
evaluating them at $\mathcal B$, and the thermal lift in the BMS
representative on the stationary horizon circle, with $x$, $y$ and $q$ as
in Eq.~\eqref{eq:adsl-xy-conventions}.

\subsection{BMS data and the obstruction at \texorpdfstring{$\mathcal B$}{B}}
The angular components of the BMS fields at $r=r_+$ read
\begin{equation}
 \omega^a{}_{\phi}
 =\left(-\frac b2,1,0\right),
 \qquad
 e^a{}_{\phi}
 =\left(-\frac J2,0,r_+\right),
 \qquad
 A^a{}_{\phi}
 =\left(\frac b2+\frac{r_-^2}{2\ell^2},0,0\right),
 \label{eq:adsl-field-data}
\end{equation}
with $M=x^2+y^2$ and $J=2\ell xy$. These data do not live on a bifurcation
circle. They live on the stationary circle
$\mathcal C_H=\{u=\mathrm{const},\,r=r_+\}$ of the past horizon: the
coordinate $u=t-r_*$ is retarded, so at fixed $u$ the limit $r\to r_+$ is
$t\to-\infty$.

The same BMS solution has, on $\mathcal C_H$,
\begin{equation}
 \mathring{\omega}_u=\frac1\ell(-xy,0,x),\qquad \omega_u=0,\qquad
 A_\phi=\left(\frac{y^2-q^2}{2},0,0\right),
\end{equation}
so that, for $\xi=\partial_u+\Omega_H\partial_\phi$,
\begin{equation}
 \iota_\xi K=-\iota_\xi A=\left(\frac{y(q^2+2x^2-y^2)}{2\ell x},0,-\frac{x}{\ell}\right),
 \qquad
 (\iota_\xi K)^2=\frac{x^2}{\ell^2}>0 .
 \label{eq:bms-bifurcation-obstruction}
\end{equation}
The scalar $(\iota_\xi K)^2$ is invariant under Lorentz frame changes and
constant along the horizon generators. For a contorsion tensor that is
continuous at $\mathcal B$, where $\xi=0$, the same scalar would instead tend
to zero along each generator. No Lorentz frame change therefore extends the
BMS geometric fields smoothly to $\mathcal B$. The transformation of
Appendix~\ref{app:regular-equivalence}, which acts on the independent
connection alone, changes the contorsion and removes this obstruction.

\subsection{Lorentz-only failure on \texorpdfstring{$\mathcal C_H$}{C-H}}

On $\mathcal C_H$ the Lorentz-only lift does not preserve the
full connection. Since $D_{\mathcal A}(\iota_\xi\mathcal A)=0$ on these data
(see below), a lift $\Lambda=\rho_\Lambda^aJ_a$
preserves $\mathcal A$ only if $D_{\mathcal A}\Lambda=0$. Its angular $J$ and $P$
components require $\rho_\Lambda\parallel\omega_\phi$ and
$\rho_\Lambda\parallel e_\phi$, and these two vectors in
Eq.~\eqref{eq:adsl-field-data} are not parallel, so $\rho_\Lambda=0$. In
particular the frame-determined parameter $-\kappa_{\rm sg}\widetilde n^a$, with
$\widetilde n^a\propto e^a{}_\phi$, is excluded; the $Z$ component would in
addition require $\rho_\Lambda\parallel A_\phi$, while the on-shell contorsion
$K^a=-A^a$ satisfies
\begin{equation}
 K^a{}_\phi\not\parallel e^a{}_\phi\qquad\text{on }\mathcal C_H .
 \label{eq:lorentz-only-obstruction}
\end{equation}
Under a change of representative the lift transforms as in
Eq.~\eqref{eq:thermal-lift-finite-law}. For a Lorentz $g$ both terms of
$\Lambda^g$ are Lorentz-valued, so whether a Lorentz-valued lift exists does
not depend on the Lorentz frame. In the stationary frame the only
Lorentz-valued lift is $\Lambda=0$, i.e.\ $\Psi=\iota_\xi\mathcal A$, whose
$\widehat J$ component $\iota_\xi\omega=\Omega_H\omega_\phi$ has norm
$qy/(\ell x)\neq\kappa_{\rm sg}$ and fails the thermal
normalization~\eqref{eq:adsl-thermal-norms}.

\subsection{Full compensator on \texorpdfstring{$\mathcal C_H$}{C-H}}
For the stationary data of Eq.~\eqref{eq:adsl-field-data},
$\iota_\xi\mathcal A\neq0$ on $\mathcal C_H$, and $\Psi$ and $\Lambda$ no
longer agree. On these flat solutions the exact-symmetry condition is
\begin{equation}
 D_{\mathcal A}\Psi=0,\qquad \Psi=\Lambda+\iota_\xi\mathcal A,
\end{equation}
pulled back to $\mathcal C_H$. In the BMS gauge, flatness and stationarity
together give $D_{\mathcal A}(\iota_\xi\mathcal A)=\mathcal{L}_\xi\mathcal A-\iota_\xi\mathcal F=0$,
so $\iota_\xi\mathcal A$ is covariantly constant. The thermal lift is
\begin{equation}
 \Psi=\Lambda_{\rm stat}+\iota_\xi\mathcal A,\qquad
 \Lambda_{\rm stat}=\mu_w\lambda_w,\qquad
 \lambda_w=\omega_\phi^a(J_a-Z_a),\qquad
 \mu_w=\frac{\kappa_{\rm sg}}q-\Omega_H,
 \label{eq:bms-stationary-lift}
\end{equation}
which is Eq.~\eqref{eq:exterior-lift} for $s_0=+1$, with
$X=\omega_\phi$. The compensator $\Lambda_{\rm stat}$ vanishes only when $qy=x^2-y^2$,
and the $\widehat J$ component of $\Psi$ is $(\kappa_{\rm sg}/q)\,\omega_\phi$
rather than $\iota_\xi\omega=\Omega_H\omega_\phi$. Below, the triples
$(\rho,\tau,\zeta)$ denote the components of $\Psi$, not of the bare
$\Lambda_{\rm stat}$.

\subsection{Solving the condition and fixing the norms}
\label{sec:compensator-solution}

With $\Psi=\rho^aJ_a+\tau^aP_a+\zeta^aZ_a$ and
$\chi^a\equiv\rho^a+\zeta^a$, the algebra~\eqref{eq:adsl-algebra} gives the
three components of the angular part of $D_{\mathcal A}\Psi=0$, for
stationary $\phi$-independent data on $\mathcal C_H$,
\begin{align}
 \text{(J)}&\quad (\omega\times\rho)^a=0,
 \label{eq:adsl-leg-j}\\
 \text{(P)}&\quad (\mathring{\omega}\times\tau)^a+(\bar e\times\chi)^a=0,
 \label{eq:adsl-leg-p}\\
 \text{(Z)}&\quad
 (\omega\times\zeta)^a+(A\times\chi)^a+(\bar e\times\tau)^a=0,
 \label{eq:adsl-leg-z}
\end{align}
where $\omega^a$ here denotes $\omega^a{}_\phi$, similarly for the
remaining one-forms, and $\mathring{\omega}=\omega+A$ is the Levi-Civita connection.
Frame indices are suppressed below.

The system decouples into the factors of Eq.~\eqref{eq:adsl-factor-generators}, with $\mathcal A_0=\omega$, $\mathcal A_\pm=\mathring{\omega}\pm\bar e$ and $\Psi_0=\rho$, $\Psi_\pm=\chi\pm\tau$. We fix the gravitational-sector parameters by transporting the geometric Lorentz lift from $\mathcal B$. In the stationary BMS frame this gives
\begin{equation}
 \Psi_\pm=\iota_\xi\mathcal A_\pm
 =\frac{\pm x+y}{\ell x}\mathcal A_{\pm\phi}.
\end{equation}
Additional covariantly constant (multiples of $\mathcal A_{\pm\phi}$) parameters define different lifted symmetries and are excluded by this prescription.

The $\pm$ norms follow from the horizon geometry. In the stationary frame $\mathcal{L}_\xi e^a=0$ and $\mathring{\omega}$ is torsion-free, so Cartan's identity gives $\mathring D\xi^a=(\iota_\xi\mathring{\omega}\times e)^a$ with $\xi^a=\iota_\xi e^a$: the vector $\iota_\xi\mathring{\omega}^a$ is the internal dual of $\nabla_\mu\xi_\nu$, and on the horizon $-\tfrac12\nabla_\mu\xi_\nu\nabla^\mu\xi^\nu=\kappa_{\rm sg}^2$ gives
$\|\iota_\xi\mathring{\omega}\|=\kappa_{\rm sg}$. On the horizon $\xi^a$ is null, and it is
orthogonal to $\iota_\xi\mathring{\omega}$ because $\xi$ is twist-free there,
$\xi_{[\rho}\nabla_\mu\xi_{\nu]}=0$. Hence
$\|\iota_\xi\mathcal A_\pm\|^2=\|\iota_\xi\mathring{\omega}\pm\xi/\ell\|^2=\kappa_{\rm sg}^2$.
Explicitly, $\iota_\xi\mathcal A_\pm=\tfrac{\pm x+y}{\ell x}\,\mathcal A_{\pm\phi}$
with $\|\mathcal A_{\pm\phi}\|=x\mp y$. This is the counterpart on
$\mathcal C_H$ of the input $\nabla^a\xi^b|_{\mathcal B}=\kappa_{\rm sg}n^{ab}$
used at a regular bifurcation circle.

In the $\widehat J$ factor, by contrast, the $J$ leg gives only
\begin{equation}
 \rho=N\,\omega_\phi ,
 \label{eq:rho-N-ansatz}
\end{equation}
with $N$ free. The free direction is the torsional gauge parameter
$\lambda_w$ of Eq.~\eqref{eq:bms-stationary-lift}, whose conserved charge is
the charge $w$ of Ref.~\cite{AvilesHidalgoValdivia2023}; adding
$c_w\lambda_w$ shifts $\rho\to\rho+c_w\omega_\phi$ and
$\zeta\to\zeta-c_w\omega_\phi$ and leaves $\chi$ and $\tau$ unchanged. The
field equations therefore do not fix $N$. The thermal
normalization~\eqref{eq:adsl-thermal-norms}, transported to this
representative by conjugation, gives $N=\kappa_{\rm sg}/q$, which fixes
$\mu_w$ in Eq.~\eqref{eq:bms-stationary-lift}. For the $\pm$ factors it is
automatic.

The signs $s_+=+1$ and $s_-=-1$ in Eq.~\eqref{eq:adsl-normalized-compensator}
below are not chosen: they are the signs of $\pm x+y$ in
$\iota_\xi\mathcal A_\pm$ above. The sign $s_0=+1$ is the branch of
Sec.~\ref{sec:adsl-lift}. With $\|\omega_\phi\|=q$ and
$\gamma\equiv\kappa_{\rm sg}/(x^2-y^2)$, the components of $\Psi$ are
\begin{equation}
 \rho=\frac{\kappa_{\rm sg}}{q}\,\omega_\phi=\kappa_{\rm sg}\left(\frac q2,\frac1q,0\right),
 \label{eq:adsl-rho}
\end{equation}
\begin{equation}
 \chi=\gamma\left(\frac{y^3-2x^2y}{2},\,y,\,x^2\right),
 \qquad
 \tau=\gamma\left(-\frac{xy^2}{2},\,x,\,xy\right),
 \label{eq:adsl-physical-compensator}
\end{equation}
and the $Z$-sector component is
\begin{equation}
 \zeta=\chi-\rho
 =\left(
 \frac{\gamma(y^3-2x^2y)-\kappa_{\rm sg}q}{2},\;
 \gamma y-\frac{\kappa_{\rm sg}}{q},\;
 \gamma x^2
 \right).
 \label{eq:adsl-varsigma-def}
\end{equation}

\subsection{Decoupling and explicit components}
\label{app:adsl-component-evaluation}

The $Z$ leg simplifies once the $J$ leg is imposed. Expanding
$(\omega+A)\times(\rho+\zeta)$ and using
Eq.~\eqref{eq:adsl-leg-j} to remove $\omega\times\rho$ leaves the first
two terms of Eq.~\eqref{eq:adsl-leg-z}, so that
Eq.~\eqref{eq:adsl-leg-z} may be replaced by
\begin{equation}
 (\mathring{\omega}\times\chi)+(\bar e\times\tau)=0
 \label{eq:adsl-leg-z-reduced}
\end{equation}
whenever the $J$ leg holds.
Taking the sum and difference of \eqref{eq:adsl-leg-p} and
\eqref{eq:adsl-leg-z-reduced}, the system
\eqref{eq:adsl-leg-j}-\eqref{eq:adsl-leg-z} is therefore equivalent to
\begin{equation}
 \omega\times\rho=0,
 \qquad
 \mathcal A_\pm\times\Psi_\pm=0,
 \qquad
 \mathcal A_\pm\equiv\mathring{\omega}\pm\bar e,
 \quad
 \Psi_\pm\equiv\chi\pm\tau,
 \label{eq:adsl-decoupled-system}
\end{equation}
the converse following by taking sums and differences again. Here
$\rho=\Psi_0$ and $\Psi_\pm$ are the components of $\Psi$ in the three
commuting factors.

Pulled back to $\mathcal C_H$, the gauge fields of
Eq.~\eqref{eq:adsl-field-data} are, in terms of $x$, $y$ and $q$,
\begin{equation}
 \omega^a{}_{\phi}
 =\left(\frac{q^2}{2},1,0\right),
 \quad
 A^a{}_{\phi}
 =\left(\frac{y^2-q^2}{2},0,0\right),
 \quad
 \bar e^a{}_{\phi}=(-xy,0,x),
 \label{eq:adsl-horizon-pullbacks}
\end{equation}
with
\begin{equation}
 \mathring{\omega}^a{}_{\phi}=\omega^a{}_\phi+A^a{}_\phi
 =\left(\frac{y^2}{2},1,0\right),
 \label{eq:adsl-shifted-pullback}
\end{equation}
and
\begin{equation}
 \mathcal A_+{}^a{}_\phi
 =\left(\frac{y^2}{2}-xy,\,1,\,x\right),
 \qquad
 \mathcal A_-{}^a{}_\phi
 =\left(\frac{y^2}{2}+xy,\,1,\,-x\right).
 \label{eq:adsl-Apm-pullback}
\end{equation}
The norms and contractions needed below are
\begin{equation}
 \omega^2=q^2,
 \qquad
 \mathring{\omega}^2=y^2,
 \qquad
 \bar e^2=x^2,
 \qquad
 \mathring{\omega}\cdot\bar e=-xy,
 \qquad
 \|\mathcal A_\pm\|^2=(x\mp y)^2,
 \label{eq:adsl-horizon-norms}
\end{equation}
where the last norm follows from
$\|\mathring{\omega}\pm\bar e\|^2=\mathring{\omega}^2\pm2\,\mathring{\omega}\cdot\bar e+\bar e^2=y^2\mp2xy+x^2$. We also note
that $A_\phi$ is null and orthogonal to $\bar e_\phi$, $A^2=A\cdot\bar e=0$.

By Eq.~\eqref{eq:centraliser} each leg is parallel to the corresponding
connection component, and the thermal normalization fixes its length;
here $\Psi_\pm=\iota_\xi\mathcal A_\pm$, whose normalization follows from the
horizon geometry, so that only the $\widehat J$ normalization is imposed:
\begin{equation}
 \rho=\kappa_{\rm sg}s_0\,\frac{\omega_\phi}{\|\omega_\phi\|},
 \qquad
 \Psi_\pm=\kappa_{\rm sg}s_\pm\,
 \frac{\mathcal A_{\pm\phi}}{\|\mathcal A_{\pm\phi}\|}.
 \label{eq:adsl-normalized-compensator}
\end{equation}
The explicit components are
\begin{equation}
 \Psi_+
 =\frac{\kappa_{\rm sg}}{x-y}
 \Bigl(\frac{y^2}{2}-xy,\,1,\,x\Bigr),
 \qquad
 \Psi_-
 =-\frac{\kappa_{\rm sg}}{x+y}
 \Bigl(\frac{y^2}{2}+xy,\,1,\,-x\Bigr).
 \label{eq:adsl-explicit-lambda-pm}
\end{equation}
The $\rho$ leg is determined independently of $\Psi_\pm$.
Using $\|\omega_\phi\|=q$ in
Eq.~\eqref{eq:adsl-normalized-compensator} gives
Eq.~\eqref{eq:adsl-rho}. Inverting the remaining two legs,
\begin{equation}
 \chi=\tfrac12(\Psi_++\Psi_-),
 \qquad
 \tau=\tfrac12(\Psi_+-\Psi_-),
\end{equation}
yields the vectors in Eq.~\eqref{eq:adsl-physical-compensator}. Finally,
subtracting $\rho$ from $\chi$ gives Eq.~\eqref{eq:adsl-varsigma-def}.
In the null basis these satisfy
\begin{equation}
 \chi\cdot\chi
 =\gamma^2\bigl(x^2-y^2\bigr)^2
 =\kappa_{\rm sg}^2,
 \qquad
 \tau\cdot\tau=0,
 \qquad
 \chi\cdot\tau=0
 \label{eq:adsl-compensator-checks}
\end{equation}
These identities verify $\Psi_\pm^2=\kappa_{\rm sg}^2$. The additional condition $\tau^2=0$ follows from the null horizon generator.

\subsection{Contractions on \texorpdfstring{$\mathcal C_H$}{C-H}}
From Eqs.~\eqref{eq:adsl-horizon-pullbacks}
and~\eqref{eq:adsl-rho}-\eqref{eq:adsl-physical-compensator},
\begin{align}
 \tau\cdot\mathring{\omega}
 &=\gamma\left(-\frac{xy^2}{2}+\frac{xy^2}{2}\right)=0,
 &
 \tau\cdot\bar e
 &=\gamma(-x^2y+x^2y)=0,
 \label{eq:adsl-exact-vanishings}\\
 \chi\cdot\bar e
 &=\gamma x(x^2-y^2)=\kappa_{\rm sg}x,
 &
 \chi \cdot \mathring{\omega}&=-\gamma y(x^2-y^2)=-\kappa_{\rm sg}y,
 \notag\\
 \rho\cdot\omega
 &=\frac{\kappa_{\rm sg}q}{2}
 +\frac{\kappa_{\rm sg}q}{2}
 =\kappa_{\rm sg}q.
 \label{eq:adsl-surviving-contractions}
\end{align}
They agree with the contractions of the geometric lift with the regular
fields~\eqref{eq:adsl-smooth-pullbacks} at $\mathcal B$, where
$\rho=\chi=-\kappa_{\rm sg}\widetilde n$ and $\tau=0$. This is consistent
with the transport~\eqref{eq:thermal-lift-finite-law}: under the
$\phi$-independent transformations of Appendix~\ref{app:regular-equivalence}
both $\Psi_i$ and $\mathcal A_{i\phi}$ are conjugated, so the factor
pairings $\Psi_i\cdot\mathcal A_{i\phi}$ are unchanged, and here
$\tau\cdot\mathring{\omega}=\tau\cdot\bar e=0$ on both circles. The Lorentz part of the isolated
mixed term is different. With these data,
$\frac{2\pi}{\kappa_{\rm sg}}\oint_{\mathcal C_H}\beta A_a\rho^a
=\pi k\alpha_2(y^2-q^2)/q$, which differs from
Eq.~\eqref{eq:adsl-wz-smooth}: identifying the mixed term with the
$\alpha_2$ block depends on the representative.

\bibliographystyle{JHEP}
\bibliography{biblio_completed}

@article{WenZee1992,
     author = "Wen, X. G. and Zee, A.",
     title = "{Shift and spin vector: New topological quantum numbers for the Hall fluids}",
     journal = "Phys. Rev. Lett.",
     volume = "69",
     pages = "953",
     year = "1992",
     doi = "10.1103/PhysRevLett.69.953",
     note = "Erratum: Phys. Rev. Lett. {\bf 69} (1992) 3000"
}

@article{GolkarRobertsSon2014,
     author = "Golkar, Siavash and Roberts, Matthew M. and Son, Dam Thanh",
     title = "{Effective field theory of relativistic quantum Hall systems}",
     eprint = "1403.4279",
     archivePrefix = "arXiv",
     primaryClass = "hep-th",
     doi = "10.1007/JHEP12(2014)138",
     journal = "JHEP",
     volume = "12",
     pages = "138",
     year = "2014"
}

@article{GolkarRobertsSonEuler2015,
     author = "Golkar, Siavash and Roberts, Matthew M. and Son, Dam Thanh",
     title = "{The Euler current and relativistic parity odd transport}",
     eprint = "1407.7540",
     archivePrefix = "arXiv",
     primaryClass = "hep-th",
     journal = "JHEP",
     volume = "04",
     pages = "110",
     year = "2015",
     doi = "10.1007/JHEP04(2015)110"
}

@article{AbanovGromov2014,
     author = "Abanov, A. G. and Gromov, A.",
     title = "{Electromagnetic and gravitational responses of two-dimensional noninteracting electrons in a background magnetic field}",
     eprint = "1401.3703",
     archivePrefix = "arXiv",
     doi = "10.1103/PhysRevB.90.014435",
     journal = "Phys. Rev. B",
     volume = "90",
     pages = "014435",
     year = "2014"
}

@article{GromovFramingAnomaly2015,
     author = "Gromov, A. and Cho, G. Y. and You, Y. and Abanov, A. G. and Fradkin, E.",
     title = "{Framing anomaly in the effective theory of the fractional quantum Hall effect}",
     eprint = "1410.6812",
     archivePrefix = "arXiv",
     doi = "10.1103/PhysRevLett.114.016805",
     journal = "Phys. Rev. Lett.",
     volume = "114",
     pages = "016805",
     year = "2015",
     note = "Erratum: Phys. Rev. Lett. {\bf 114} (2015) 149902"
}

@article{CappelliRandellini2016,
     author = "Cappelli, A. and Randellini, E.",
     title = "{Multipole expansion in the quantum Hall effect}",
     eprint = "1512.02147",
     archivePrefix = "arXiv",
     doi = "10.1007/JHEP03(2016)105",
     journal = "JHEP",
     volume = "03",
     pages = "105",
     year = "2016"
}

@article{BacryCombeRichard1970,
     author = "Bacry, H. and Combe, P. and Richard, J. L.",
     title = "{Group-theoretical analysis of elementary particles in an external electromagnetic field. 1. The relativistic particle in a constant and uniform field}",
     journal = "Nuovo Cim. A",
     volume = "67",
     pages = "267",
     year = "1970",
     doi = "10.1007/BF02725178"
}

@article{Schrader1972,
     author = "Schrader, R.",
     title = "{The Maxwell group and the quantum theory of particles in classical homogeneous electromagnetic fields}",
     journal = "Fortsch. Phys.",
     volume = "20",
     pages = "701",
     year = "1972",
     doi = "10.1002/prop.19720201202"
}

@article{Palumbo2016,
     author = "Palumbo, G.",
     title = "{Geometric model of topological insulators from the Maxwell algebra}",
     eprint = "1610.04734",
     archivePrefix = "arXiv",
     primaryClass = "cond-mat.mes-hall",
     doi = "10.1016/j.aop.2017.08.018",
     journal = "Annals Phys.",
     volume = "386",
     pages = "15",
     year = "2017"
}

@article{DurkaKG2019,
     author = "Durka, R. and Kowalski-Glikman, J.",
     title = "{Resonant algebras in Chern-Simons model of topological insulators}",
     eprint = "1906.02356",
     archivePrefix = "arXiv",
     primaryClass = "hep-th",
     doi = "10.1016/j.physletb.2019.06.058",
     journal = "Phys. Lett. B",
     volume = "795",
     pages = "516",
     year = "2019"
}

@article{Concha2018,
     author = "Concha, P. and Merino, N. and Miskovic, O. and Rodriguez, E. and Salgado-Rebolledo, P. and Valdivia, O.",
     title = "{Asymptotic symmetries of three-dimensional Chern-Simons gravity for the Maxwell algebra}",
     eprint = "1805.08834",
     archivePrefix = "arXiv",
     doi = "10.1007/JHEP10(2018)079",
     journal = "JHEP",
     volume = "10",
     pages = "079",
     year = "2018"
}

@article{Aviles2026,
     author = "Avil\'es, L. and Concha, P. and Matulich, J. and Rodr\'iguez, E. and Tempo, D.",
     title = "{Maxwell Chern-Simons gravity in 3D: thermodynamics of cosmological solutions and black holes with torsion}",
     eprint = "2510.25559",
     archivePrefix = "arXiv",
     primaryClass = "hep-th",
     doi = "10.1007/JHEP02(2026)143",
     journal = "JHEP",
     volume = "02",
     pages = "143",
     year = "2026"
}

@article{AvilesHidalgoValdivia2023,
     author = "Avil\'es, L. and Hidalgo, D. and Valdivia, O.",
     title = "{Thermodynamics of the three-dimensional black hole with torsion}",
     eprint = "2308.09114",
     archivePrefix = "arXiv",
     primaryClass = "hep-th",
     doi = "10.1007/JHEP09(2023)185",
     journal = "JHEP",
     volume = "09",
     pages = "185",
     year = "2023"
}

@article{Witten1988,
     author = "Witten, E.",
     title = "{(2+1)-dimensional gravity as an exactly soluble system}",
     journal = "Nucl. Phys. B",
     volume = "311",
     pages = "46",
     year = "1988",
     doi = "10.1016/0550-3213(88)90143-5"
}

@article{Zanelli2012,
     author = "Zanelli, J.",
     title = "{Chern-Simons forms in gravitation theories}",
     eprint = "1208.3353",
     archivePrefix = "arXiv",
     doi = "10.1088/0264-9381/29/13/133001",
     journal = "Class. Quant. Grav.",
     volume = "29",
     pages = "133001",
     year = "2012"
}

@article{Wald1993,
     author = "Wald, R. M.",
     title = "{Black hole entropy is the Noether charge}",
     eprint = "gr-qc/9307038",
     archivePrefix = "arXiv",
     doi = "10.1103/PhysRevD.48.R3427",
     journal = "Phys. Rev. D",
     volume = "48",
     pages = "R3427",
     year = "1993"
}

@article{IyerWald1994,
     author = "Iyer, V. and Wald, R. M.",
     title = "{Some properties of Noether charge and a proposal for dynamical black hole entropy}",
     eprint = "gr-qc/9403028",
     archivePrefix = "arXiv",
     doi = "10.1103/PhysRevD.50.846",
     journal = "Phys. Rev. D",
     volume = "50",
     pages = "846",
     year = "1994"
}

@article{JacobsonMohd,
     author = "Jacobson, T. and Mohd, A.",
     title = "{Black hole entropy and Lorentz-diffeomorphism Noether charge}",
     eprint = "1507.01054",
     archivePrefix = "arXiv",
     primaryClass = "gr-qc",
     doi = "10.1103/PhysRevD.92.124010",
     journal = "Phys. Rev. D",
     volume = "92",
     number = "12",
     pages = "124010",
     year = "2015"
}

@article{Tachikawa2007,
     author = "Tachikawa, Y.",
     title = "{Black hole entropy in the presence of Chern-Simons terms}",
     eprint = "hep-th/0611141",
     archivePrefix = "arXiv",
     doi = "10.1088/0264-9381/24/3/014",
     journal = "Class. Quant. Grav.",
     volume = "24",
     pages = "737",
     year = "2007"
}

@article{SalgadoRebolledoPalumbo2021,
     author = "Salgado-Rebolledo, Patricio and Palumbo, Giandomenico",
     title = "{Extended Nappi-Witten geometry for the fractional quantum Hall effect}",
     eprint = "2102.03886",
     archivePrefix = "arXiv",
     doi = "10.1103/PhysRevD.103.125006",
     journal = "Phys. Rev. D",
     volume = "103",
     number = "12",
     pages = "125006",
     year = "2021"
}

@article{BradlynRead2015,
  author  = {Bradlyn, Barry and Read, N.},
  title   = {Low-energy effective theory in the bulk for transport in a topological phase},
  journal = {Phys. Rev. B},
  volume  = {91},
  pages   = {125303},
  year    = {2015},
  doi     = {10.1103/PhysRevB.91.125303},
  eprint  = {1407.2911},
  archivePrefix = {arXiv},
  primaryClass  = {cond-mat.str-el}
}

@article{Banados1995GlobalCharges,
  author = {Ba\~nados, Maximo},
  title = {{Global charges in Chern-Simons theory and the 2+1 black hole}},
  journal = {Phys. Rev. D},
  volume = {52},
  pages = {5816},
  year = {1995},
  doi = {10.1103/PhysRevD.52.5816},
  eprint = {hep-th/9405171},
  archivePrefix = {arXiv}
}

@article{Witten1989,
     author = "Witten, E.",
     title = "{Quantum field theory and the Jones polynomial}",
     journal = "Commun. Math. Phys.",
     volume = "121",
     pages = "351",
     year = "1989",
     doi = "10.1007/BF01217730"
}

@article{Wen1995,
     author = "Wen, X. G.",
     title = "{Topological orders and edge excitations in fractional quantum Hall states}",
     eprint = "cond-mat/9506066",
     archivePrefix = "arXiv",
     journal = "Adv. Phys.",
     volume = "44",
     pages = "405",
     year = "1995",
     doi = "10.1080/00018739500101566"
}

@article{Read2009,
     author = "Read, N.",
     title = "{Non-Abelian adiabatic statistics and Hall viscosity in quantum Hall states and $p_x+ip_y$ paired superfluids}",
     eprint = "0805.2507",
     archivePrefix = "arXiv",
     journal = "Phys. Rev. B",
     volume = "79",
     pages = "045308",
     year = "2009",
     doi = "10.1103/PhysRevB.79.045308"
}

@article{AchucarroTownsend1986,
     author = "Ach\'ucarro, A. and Townsend, P. K.",
     title = "{A Chern-Simons action for three-dimensional anti-de Sitter supergravity theories}",
     journal = "Phys. Lett. B",
     volume = "180",
     pages = "89",
     year = "1986",
     doi = "10.1016/0370-2693(86)90140-1"
}

@article{BTZ1992,
     author = "Ba\~nados, M. and Teitelboim, C. and Zanelli, J.",
     title = "{The black hole in three-dimensional spacetime}",
     eprint = "hep-th/9204099",
     archivePrefix = "arXiv",
     journal = "Phys. Rev. Lett.",
     volume = "69",
     pages = "1849",
     year = "1992",
     doi = "10.1103/PhysRevLett.69.1849"
}

@article{Bekenstein1973,
     author = "Bekenstein, J. D.",
     title = "{Black holes and entropy}",
     journal = "Phys. Rev. D",
     volume = "7",
     pages = "2333",
     year = "1973",
     doi = "10.1103/PhysRevD.7.2333"
}

@article{Hawking1975,
     author = "Hawking, S. W.",
     title = "{Particle creation by black holes}",
     journal = "Commun. Math. Phys.",
     volume = "43",
     pages = "199",
     year = "1975",
     doi = "10.1007/BF02345020"
}

@article{Ammon2013,
    author = "Ammon, Martin and Gutperle, Michael and Kraus, Per and Perlmutter, Eric",
    title = "{Black holes in three dimensional higher spin gravity: A review}",
    eprint = "1208.5182",
    archivePrefix = "arXiv",
    primaryClass = "hep-th",
    doi = "10.1088/1751-8113/46/21/214001",
    journal = "J. Phys. A",
    volume = "46",
    pages = "214001",
    year = "2013"
}

@article{Anandan1996,
    author = "Anandan, J.",
    title = "{Gravitational phase operator and cosmic strings}",
    eprint = "gr-qc/9507049",
    archivePrefix = "arXiv",
    doi = "10.1103/PhysRevD.53.779",
    journal = "Phys. Rev. D",
    volume = "53",
    pages = "779--786",
    year = "1996"
}

@article{BTZ1994,
    author = "Banados, Maximo and Teitelboim, Claudio and Zanelli, Jorge",
    title = "{Black hole entropy and the dimensional continuation of the Gauss-Bonnet theorem}",
    eprint = "gr-qc/9309026",
    archivePrefix = "arXiv",
    reportNumber = "IASSNS-HEP-93-53",
    doi = "10.1103/PhysRevLett.72.957",
    journal = "Phys. Rev. Lett.",
    volume = "72",
    pages = "957--960",
    year = "1994"
}

@article{Birmingham1991,
    author = "Birmingham, Danny and Blau, Matthias and Rakowski, Mark and Thompson, George",
    title = "{Topological field theory}",
    reportNumber = "CERN-TH-6045-91",
    doi = "10.1016/0370-1573(91)90117-5",
    journal = "Phys. Rept.",
    volume = "209",
    pages = "129--340",
    year = "1991"
}

@article{Bradlyn2015,
    author = "Bradlyn, Barry and Read, N.",
    title = "{Topological central charge from Berry curvature: Gravitational anomalies in trial wave functions for topological phases}",
    eprint = "1502.04126",
    archivePrefix = "arXiv",
    primaryClass = "cond-mat.mes-hall",
    doi = "10.1103/PhysRevB.91.165306",
    journal = "Phys. Rev. B",
    volume = "91",
    number = "16",
    pages = "165306",
    year = "2015"
}

@article{BunsterEtAl2014,
    author = "Bunster, Claudio and Henneaux, Marc and Perez, Alfredo and Tempo, David and Troncoso, Ricardo",
    title = "{Generalized Black Holes in Three-dimensional Spacetime}",
    eprint = "1404.3305",
    archivePrefix = "arXiv",
    primaryClass = "hep-th",
    doi = "10.1007/JHEP05(2014)031",
    journal = "JHEP",
    volume = "05",
    pages = "031",
    year = "2014"
}

@article{Can2014,
    author = "Can, T. and Laskin, M. and Wiegmann, P.",
    title = "{Fractional Quantum Hall Effect in a Curved Space: Gravitational Anomaly and Electromagnetic Response}",
    eprint = "1402.1531",
    archivePrefix = "arXiv",
    primaryClass = "cond-mat.str-el",
    doi = "10.1103/PhysRevLett.113.046803",
    journal = "Phys. Rev. Lett.",
    volume = "113",
    pages = "046803",
    year = "2014"
}

@article{Can2016,
    author = "Can, T. and Chiu, Y. H. and Laskin, M. and Wiegmann, P.",
    title = "{Emergent Conformal Symmetry and Geometric Transport Properties of Quantum Hall States on Singular Surfaces}",
    eprint = "1602.04802",
    archivePrefix = "arXiv",
    primaryClass = "cond-mat.str-el",
    doi = "10.1103/PhysRevLett.117.266803",
    journal = "Phys. Rev. Lett.",
    volume = "117",
    number = "26",
    pages = "266803",
    year = "2016"
}

@article{CarlipTeitelboim1995,
    author = "Carlip, Steven and Teitelboim, Claudio",
    title = "{The Off-shell black hole}",
    eprint = "gr-qc/9312002",
    archivePrefix = "arXiv",
    reportNumber = "IASSNS-HEP-93-84, UCD-93-34",
    doi = "10.1088/0264-9381/12/7/011",
    journal = "Class. Quant. Grav.",
    volume = "12",
    pages = "1699--1704",
    year = "1995"
}

@incollection{Dunne1999,
    author = "Dunne, Gerald V.",
    title = "{Aspects of Chern-Simons Theory}",
    booktitle = "{Topological Aspects of Low Dimensional Systems}",
    series = "Les Houches",
    volume = "69",
    publisher = "Springer",
    pages = "177--263",
    year = "1999",
    doi = "10.1007/3-540-46637-1_3",
    eprint = "hep-th/9902115",
    archivePrefix = "arXiv"
}

@article{Fradkin1998,
    author = "Fradkin, Eduardo H. and Nayak, Chetan and Tsvelik, A. and Wilczek, Frank",
    title = "{A Chern-Simons effective field theory for the Pfaffian quantum Hall state}",
    eprint = "cond-mat/9711087",
    archivePrefix = "arXiv",
    reportNumber = "NSF-ITP-97-134",
    doi = "10.1016/S0550-3213(98)00111-4",
    journal = "Nucl. Phys. B",
    volume = "516",
    pages = "704--718",
    year = "1998"
}

@article{FursaevSolodukhin1995,
    author = "Fursaev, Dmitri V. and Solodukhin, Sergey N.",
    title = "{On the description of the Riemannian geometry in the presence of conical defects}",
    eprint = "hep-th/9501127",
    archivePrefix = "arXiv",
    reportNumber = "JINR-E2-95-28, JINR, E2-95-28",
    doi = "10.1103/PhysRevD.52.2133",
    journal = "Phys. Rev. D",
    volume = "52",
    pages = "2133--2143",
    year = "1995"
}

@article{GusyninSharapov2005,
    author = "Gusynin, V. P. and Sharapov, S. G.",
    title = "{Unconventional integer quantum Hall effect in graphene}",
    eprint = "cond-mat/0506575",
    archivePrefix = "arXiv",
    doi = "10.1103/PhysRevLett.95.146801",
    journal = "Phys. Rev. Lett.",
    volume = "95",
    pages = "146801",
    year = "2005"
}

@article{Han2019,
    author = "Han, Bo and Wang, Huajia and Ye, Peng",
    title = "{Generalized Wen-Zee Terms}",
    eprint = "1807.10844",
    archivePrefix = "arXiv",
    primaryClass = "cond-mat.str-el",
    doi = "10.1103/PhysRevB.99.205120",
    journal = "Phys. Rev. B",
    volume = "99",
    number = "20",
    pages = "205120",
    year = "2019"
}

@article{HorneWitten1989,
    author = "Horne, James H. and Witten, Edward",
    title = "{Conformal Gravity in Three-dimensions as a Gauge Theory}",
    reportNumber = "PUPT-1109",
    doi = "10.1103/PhysRevLett.62.501",
    journal = "Phys. Rev. Lett.",
    volume = "62",
    pages = "501--504",
    year = "1989"
}

@article{Hoyos2012,
    author = "Hoyos, Carlos and Son, Dam Thanh",
    title = "{Hall Viscosity and Electromagnetic Response}",
    eprint = "1109.2651",
    archivePrefix = "arXiv",
    primaryClass = "cond-mat.mes-hall",
    reportNumber = "INT-PUB-11-037",
    doi = "10.1103/PhysRevLett.108.066805",
    journal = "Phys. Rev. Lett.",
    volume = "108",
    pages = "066805",
    year = "2012"
}

@article{Kraus2013,
    author = "Kraus, Per and Ugajin, Tomonori",
    title = "{An Entropy Formula for Higher Spin Black Holes via Conical Singularities}",
    eprint = "1302.1583",
    archivePrefix = "arXiv",
    primaryClass = "hep-th",
    reportNumber = "IPMU13-0023, YITP-13-4",
    doi = "10.1007/JHEP05(2013)160",
    journal = "JHEP",
    volume = "05",
    pages = "160",
    year = "2013"
}

@article{Lopez1991,
    author = "Lopez, Ana and Fradkin, Eduardo",
    title = "{Fractional quantum Hall effect and Chern-Simons gauge theories}",
    doi = "10.1103/PhysRevB.44.5246",
    journal = "Phys. Rev. B",
    volume = "44",
    number = "10",
    pages = "5246--5262",
    year = "1991"
}

@article{Novoselov2005,
    author = "Novoselov, K. S. and Geim, A. K. and Morozov, S. V. and Jiang, D. and Katsnelson, M. I. and Grigorieva, I. V. and Dubonos, S. V. and Firsov, A. A.",
    title = "{Two-dimensional gas of massless Dirac fermions in graphene}",
    eprint = "cond-mat/0509330",
    archivePrefix = "arXiv",
    doi = "10.1038/nature04233",
    journal = "Nature",
    volume = "438",
    pages = "197",
    year = "2005"
}

@article{Palumbo2013,
    author = "Palumbo, Giandomenico and Pachos, Jiannis K.",
    title = "{Abelian Chern-Simons-Maxwell theory from a tight binding model of spinless fermions}",
    eprint = "1301.2625",
    archivePrefix = "arXiv",
    primaryClass = "cond-mat.str-el",
    doi = "10.1103/PhysRevLett.110.211603",
    journal = "Phys. Rev. Lett.",
    volume = "110",
    number = "21",
    pages = "211603",
    year = "2013"
}

@article{Palumbo2014,
    author = "Palumbo, Giandomenico and Pachos, Jiannis K.",
    title = "{Non-Abelian Chern-Simons Theory from a Hubbard-like Model}",
    eprint = "1311.2871",
    archivePrefix = "arXiv",
    primaryClass = "cond-mat.str-el",
    doi = "10.1103/PhysRevD.90.027703",
    journal = "Phys. Rev. D",
    volume = "90",
    number = "2",
    pages = "027703",
    year = "2014"
}

@article{Prabhu2017,
    author = "Prabhu, Kartik",
    title = "{The First Law of Black Hole Mechanics for Fields with Internal Gauge Freedom}",
    eprint = "1511.00388",
    archivePrefix = "arXiv",
    primaryClass = "gr-qc",
    doi = "10.1088/1361-6382/aa536b",
    journal = "Class. Quant. Grav.",
    volume = "34",
    number = "3",
    pages = "035011",
    year = "2017"
}

@article{Rao2023,
    author = "Rao, Pranav and Bradlyn, Barry",
    title = "{Resolving Hall and dissipative viscosity ambiguities via boundary effects}",
    eprint = "2112.04545",
    archivePrefix = "arXiv",
    primaryClass = "cond-mat.mes-hall",
    doi = "10.1103/PhysRevB.107.075148",
    journal = "Phys. Rev. B",
    volume = "107",
    number = "7",
    pages = "075148",
    year = "2023"
}

@article{Salgado2019,
    author = "Salgado-Rebolledo, Patricio",
    title = "{The Maxwell group in 2+1 dimensions and its infinite-dimensional enhancements}",
    eprint = "1905.09421",
    archivePrefix = "arXiv",
    primaryClass = "hep-th",
    doi = "10.1007/JHEP10(2019)039",
    journal = "JHEP",
    volume = "10",
    pages = "039",
    year = "2019"
}

@article{Salgado2020,
    author = "Salgado-Rebolledo, Patricio and Palumbo, Giandomenico and Pachos, Jiannis K.",
    title = "{Effective field theories for interacting boundaries of 3D topological crystalline insulators through bosonisation}",
    eprint = "2006.04752",
    archivePrefix = "arXiv",
    primaryClass = "cond-mat.mes-hall",
    doi = "10.1038/s41598-020-77966-3",
    journal = "Sci. Rep.",
    volume = "10",
    number = "1",
    pages = "21998",
    year = "2020"
}

@book{Wen2004,
    author = "Wen, Xiao-Gang",
    title = "{Quantum Field Theory of Many-Body Systems: From the Origin of Sound to an Origin of Light and Electrons}",
    publisher = "Oxford University Press",
    year = "2004",
    isbn = "9780198530947"
}

@article{Yakovenko1990,
    author = "Yakovenko, V. M.",
    title = "{Chern-Simons terms and $n$ field in Haldane's model for the quantum Hall effect without Landau levels}",
    journal = "Phys. Rev. Lett.",
    volume = "65",
    pages = "251--254",
    year = "1990",
    doi = "10.1103/PhysRevLett.65.251"
}

@article{Zhang1989,
    author = "Zhang, S. C. and Hansson, T. H. and Kivelson, S.",
    title = "{Effective-Field-Theory Model for the Fractional Quantum Hall Effect}",
    reportNumber = "NSF-ITP-88-119",
    doi = "10.1103/PhysRevLett.62.82",
    journal = "Phys. Rev. Lett.",
    volume = "62",
    pages = "82--85",
    year = "1989"
}

@article{Zhang2005,
    author = "Zhang, Yuanbo and Tan, Yan-Wen and Stormer, Horst L. and Kim, Philip",
    title = "{Experimental observation of the quantum Hall effect and Berry's phase in graphene}",
    eprint = "cond-mat/0509355",
    archivePrefix = "arXiv",
    doi = "10.1038/nature04235",
    journal = "Nature",
    volume = "438",
    pages = "201--204",
    year = "2005"
}

@article{deBoerJottar2014,
    author = "de Boer, Jan and Jottar, Juan I.",
    title = "{Thermodynamics of higher spin black holes in $AdS_3$}",
    eprint = "1302.0816",
    archivePrefix = "arXiv",
    primaryClass = "hep-th",
    doi = "10.1007/JHEP01(2014)023",
    journal = "JHEP",
    volume = "01",
    pages = "023",
    year = "2014"
}

@article{ToplogicalgravityTH,
     author = "Salgado, Patricio and Szabo, Richard J. and Valdivia, Omar",
     title = "{Topological gravity and transgression holography}",
     eprint = "1401.3653",
     archivePrefix = "arXiv",
     primaryClass = "hep-th",
     doi = "10.1103/PhysRevD.89.084077",
     journal = "Phys. Rev. D",
     volume = "89",
     number = "8",
     pages = "084077",
     year = "2014"
}

@article{Concha2019,
     author = "Concha, Patrick and Merino, Nelson and Rodr{\'\i}guez, Evelyn and Salgado-Rebolledo, Patricio and Valdivia, Omar",
     title = "{Semi-simple enlargement of the $\mathfrak{bms}_3$ algebra from a $\mathfrak{so}(2,2)\oplus\mathfrak{so}(2,1)$ Chern-Simons theory}",
     eprint = "1810.12256",
     archivePrefix = "arXiv",
     primaryClass = "hep-th",
     doi = "10.1007/JHEP02(2019)002",
     journal = "JHEP",
     volume = "02",
     pages = "002",
     year = "2019"
}

@article{Diaz2012,
     author = "D{\'\i}az, J. and Fierro, O. and Izaurieta, F. and Merino, N. and Rodr{\'\i}guez, E. and Salgado, P. and Valdivia, O.",
     title = "{A generalized action for (2+1)-dimensional Chern-Simons gravity}",
     eprint = "1311.2215",
     archivePrefix = "arXiv",
     primaryClass = "gr-qc",
     doi = "10.1088/1751-8113/45/25/255207",
     journal = "J. Phys. A",
     volume = "45",
     number = "25",
     pages = "255207",
     year = "2012"
}

@article{2025boundarydynamicsmaxwellinvariantthreedimensional,
     author = "H{\"o}fenstock, Felix and Salgado-Rebolledo, Patricio",
     title = "{Boundary dynamics of Maxwell-invariant three-dimensional Chern-Simons gravity}",
     eprint = "2506.07651",
     archivePrefix = "arXiv",
     primaryClass = "hep-th",
     doi = "10.1007/JHEP01(2026)022",
     journal = "JHEP",
     volume = "01",
     pages = "022",
     year = "2026"
}
\end{document}